\documentclass[usenatbib]{mnras}
\usepackage{graphics}
\usepackage{lipsum}
\usepackage{amsmath}
\usepackage{breqn}
\usepackage{amssymb}
\usepackage{hyperref}
\usepackage{subfigure}
\usepackage{multicol}
\usepackage{caption}
\usepackage{multirow}
\usepackage{ulem}
\usepackage{rotating}
\usepackage{pdflscape}

\usepackage{color}

\title{Multiplicity functions of quasars: Predictions from the \texttt{MassiveBlackII} simulation}
\author[Bhowmick et al.]{
Aklant K. Bhowmick$^{1}$,
Tiziana Di Matteo$^{1}$,
Adam D. Myers$^{2}$
\\
$^{1}$McWilliams Center for Cosmology, Dept. of Physics, Carnegie Mellon University,
Pittsburgh PA 15213, USA\\
$^{2}$Department of Physics \& Astronomy, University of Wyoming, 1000 University Ave., Laramie, WY, 82071\\}
\begin{document}
\maketitle
\begin{abstract}

We examine multiple AGN systems (triples and quadruples, in particular) in the \texttt{MassiveBlackII} simulation over a redshift range of $0.06\lesssim z \lesssim 4$. We identify AGN systems (with bolometric luminosity $L_{\mathrm{bol}}>10^{42}~\mathrm{ergs/sec}$) at different scales~(defined by the maximum distance between member AGNs) to determine the AGN multiplicity functions. This is defined as the volume/ surface density of AGN systems per unit \textit{richness} $R$, the number of AGNs in a system. We find that gravitationally bound multiple AGN systems tend to populate scales of $\lesssim0.7~\mathrm{cMpc}/h$; this corresponds to angular separations of $\lesssim100~\mathrm{arcsec}$ and a line of sight velocity difference $\lesssim200~\mathrm{km/sec}$. The simulation contains $\sim 10$ and $\sim100$  triples/quadruples per $\mathrm{deg}^2$ up to depths of DESI~($g\lesssim24$) and LSST~($g\lesssim26$) imaging respectively; at least $20\%$ of these should be detectable in spectroscopic surveys. The simulated quasar~($L_{\mathrm{bol}}>10^{44}~\mathrm{ergs/sec}$) triples and quadruples predominantly exist at  $1.5\lesssim z \lesssim 3$. Their members have black hole masses $10^{6.5}\lesssim M_{bh}\lesssim 10^{9}~M_{\odot}/h$ and live in separate (one central and multiple satellite) galaxies with stellar masses $10^{10}\lesssim M_{*}\lesssim 10^{12}~M_{\odot}/h$. They live in the most massive haloes 
(for e.g. $\sim 10^{13}~M_{\odot}/h$ at $z=2.5$; $\sim 10^{14}~M_{\odot}/h$ at $z=1$) in the simulation. Their detections provide an exciting
prospect for understanding massive black hole growth and their merger rates in galaxies in the era of multi-messenger astronomy.
\end{abstract}

\begin{keywords}
quasars: general,  close pairs
\end{keywords}

%\newline
%\newline

\section{Introduction}
Halo mergers are a very important component in the current paradigm of galaxy formation, and are one of the primary mechanisms via which galaxies and dark matter haloes grow and evolve. Furthermore, the ensuing interactions between the galaxies that occupy recently merged haloes can be possible triggers of AGN activity \citep{2005ApJ...621...95W,2005Natur.433..604D}. If true, this scenario should lead to multiple simultaneously active AGNs within the same galaxy/~halo. Observational signatures have been found in ``excess'' small scale~($\lesssim 1~\mathrm{Mpc}$) clustering measurements compared to large scales~ \citep{2000AJ....120.2183S,2006AJ....131....1H,2007ApJ...658...99M,2008ApJ...678..635M,2010ApJ...719.1693S,2012MNRAS.424.1363K,2016AJ....151...61M,2017MNRAS.468...77E}. These measurements mostly originate from ``binary'' quasar pairs obtained from the SDSS quasar catalogs~\citep{2010AJ....139.2360S} which are very rare objects ($1$ pair in $\sim 10^4$ quasars). Recent analysis \citep{2019MNRAS.485.2026B} of these measurements using the \texttt{MassiveBlackII} simulation \citep{2015MNRAS.450.1349K} provide further support for halo mergers as likely origins of quasars pairs.

Systems with more than two AGNs~(triples and quadruples) are believed to be even more rare and elusive. Over the past decade, two AGN triples \citep{2007ApJ...662L...1D,2013MNRAS.431.1019F,2019arXiv190710639L} at $z\sim1.5,2$ and one quadruple \citep{2015Sci...348..779H} at $z\sim2$ have been discovered. These quasar systems have associated length scales~(separation between member AGNs) of a few hundred kpcs. Fainter AGN systems have also been found at smaller scales \citep[e.g.][]{2011ApJ...743L..37S,2011ApJ...736L...7L}. These different systems were discovered somewhat independently and originate from different observational datasets. For example, the \cite{2007ApJ...662L...1D} triple was discovered using the data from the  Low-Resolution Imaging Spectrometer~\citep[LRIS]{1995PASP..107..375O}  at the W. M. Keck Observatory. The \cite{2013MNRAS.431.1019F} triplet was discovered as part of a dedicated program for search of AGN triples within a sample of SDSS quasars compiled by \cite{2009ApJS..180...67R}. The quasar quadruple in \cite{2015Sci...348..779H} was found as part of a search for extended Lyman alpha emission within a sample of 29 quasars. The faintest members of these systems have apparent magnitudes   of $\sim21$. As we shall see in this work, state of the art hydrodynamic simulations can probe quasar systems with magnitudes up to $\sim24$ at $z\sim1.5,2$. While simulated AGN systems may be somewhat fainter than the observed quasar systems, studying their properties will nevertheless provide significant insight into the physics involved in the formation of such systems.

Physical processes underlying the formation and evolution of such AGN systems remain uncertain. Analyses in \cite{2013MNRAS.431.1019F} and \cite{2015Sci...348..779H} reveal that these systems are in extreme environments in the distant ($z\gtrsim2$) universe, and are likely progenitors of present-day massive clusters; this suggests that such systems may act as signposts for tracing the early stages in the evolution of galaxy clusters, and can thereby potentially reveal new insights into galaxy evolution. 

The incidence of AGN pairs and triples and the associated host galaxies is also crucial for understanding black hole growth
and the role of mergers.
Exciting prospects exist with 
the new generation of space and ground based telescopes which shall achieve a deep, almost synoptical monitoring of the whole sky; this will be accompanied by gravitational wave observatories, such as LISA~\citep{2017PhRvD..95j3012B}, which will detect these massive black holes at various stages of their merging process. With possible simultaneous detections of gravitational and electromagnetic wave emissions, we are going to be able to directly probe a wide range of properties of  these systems and their environments, for e.g. the masses of black holes and associated host galaxies and haloes.

%\tiziana{you need to say somethingmore exciting about why pairs and triples etc.. are exciting for understanding BH growth/ galaxy connection in the context also of multi-messenger astonomy, the advent of LISA. e.g. something along the lines of what I have below - but please improve somewhat} 
It is therefore crucial to investigate what current theories of structure formation coupled with galaxy evolution and black hole growth can tell us about the origin and abundance of such systems. To these ends, in this work, we study AGN systems in the \texttt{MassiveBlackII}~(hereafter MBII) cosmological hydrodynamic simulation. We present our basic methodology and simulation details in Section \ref{methods_sec}. In Section \ref{MF_sec}, we quantify the abundance of AGN systems by determining their multiplicity functions. In Section \ref{selected_systems_sec}, we select AGN systems in MBII and study the basic properties of their environments and their growth histories. In Section \ref{conclusions}, we discuss the main conclusions and possible future directions. 
%We shall be using units of 'cMpc' for \textit{comoving} distance in mega-parsecs, '$\mathrm{ergs/sec}$' for luminosities, and '$M_{\odot}/h$' for black hole and halo masses unless stated otherwise. \tiziana{do you really need to say this about the units - you have to write them anyway and they are obvious}      
\section{Methods}
\label{methods_sec}
\subsection{\texttt{MassiveBlackII}~(MBII) simulation}
\texttt{MBII} is a high-resolution cosmological hydrodynamic simulation with a box size of $100~\mathrm{cMpc/h}$ and $2\times1792^3$ particles. It runs from $z=159$ to $z=0.06$. Cosmological parameters are constrained by WMAP7 \citep{2011ApJS..192...18K} i.e. $\Omega_0=0.275$, $\Omega_l=0.725$, $\Omega_b=0.046$, $\sigma_8=0.816$, $h = 0.701$, $n_s=0.968$. The dark matter and gas particle masses are $1.1\times 10^7~M_{\odot}/h$ and $2.2\times 10^6~M_{\odot}/h$ respectively. The simulation includes a variety of subgrid physics modeling such as star formation \citep{2003MNRAS.339..289S}, black hole growth and associated feedback. Haloes and subhaloes were identified using a Friends-of-Friends (FOF) group finder \citep{1985ApJ...292..371D} and \texttt{SUBFIND} \citep{2005MNRAS.364.1105S} respectively. For more details, we refer the reader to \cite{2015MNRAS.450.1349K}.

\subsubsection{Modeling black holes in MBII}
We use the prescription described in \cite{2005Natur.433..604D} and \cite{2005MNRAS.361..776S} to model the growth of supermassive black holes. Black holes of mass $5\times 10^{5}~M_{\odot}/h$ are initially seeded into $\gtrsim 5\times 10^{10}~M_{\odot}/h$ haloes which do not already contain a black hole. The black holes then grow via accretion at a rate given by $\dot{M}_{bh}={4\pi G^2 M_{bh}^2 \rho}/{(c_s^2+v_{bh}^2)^{3/2}}$; $\rho$ and $c_s$ are the density and sound speed of the cold phase of the ISM gas, and $v_{bh}$ is the relative velocity of the black hole w.r.t gas. The accreting black holes radiate with bolometric luminosity $\epsilon_r mc^2$ where the radiative efficiency ($\epsilon_r$) is taken to be $10\%$. We allow for the accretion rate to be mildly super-Eddington but limit it to two times the Eddington accretion rate. $5\%$ of the radiated energy is thermodynamically (and isotropically) coupled to the surrounding gas as blackhole~(or AGN) feedback \citep{2005Natur.433..604D}. Additionally, black holes can can grow via merging; two black holes are considered to be merged if their separation distance is smaller than the spatial resolution of the simulation~(the SPH smoothing length), and they have a relative speed smaller than the local sound speed of the medium. Further details on black hole modeling have been described in \cite{2012ApJ...745L..29D}.

\subsubsection{Calculating g-band magnitudes of simulated AGNs}
We shall be presenting the multiplicity functions for AGN samples limited by the $g$ band apparent magnitude. To compute the apparent magnitude, we first compute the absolute magnitude $M_g~(z=2)$ from the bolometric luminosity $L_{\mathrm{bol}}$ using the following bolometric correction~\citep{2009ApJ...697.1656S,2009MNRAS.399.1755C},
\begin{eqnarray}
    M_i(z=2)=90-2.5\log_{10}\frac{L_{\mathrm{bol}}}{{\mathrm{ergs~sec^{-1}}}}\\
    M_g(z=2)=M_i(z=2)+2.5\alpha_{\nu}\log_{10}{\frac{4670~\mathrm{\AA}}{7471~\mathrm{\AA}}}
\end{eqnarray}
where $\alpha_{\nu}=-0.5$. The $g$ band apparent magnitudes are then computed using Eq.~(4) of \cite{2016A&A...587A..41P}
\begin{equation}
    M_g(z=2)=g-d_m(z)-(K(z)-K(z=2))
\end{equation} 
where $d_m(z)$ is the distance modulus and $K(z)$ is the k-correction adopted from \cite{2013ApJ...768..105M}.
\subsubsection{Identifying AGNs: Centrals and Satellites}
\label{central_satellite_identification}
AGNs are identified to be individual black holes which are accreting gas in their vicinity. In this work, we consider objects with bolometric luminosity $L_{\mathrm{bol}}\gtrsim 3\times10^{41} \mathrm{ergs/sec}$ and $M_{bh}\gtrsim 10^{6}~M_{\odot}/h$ (2 times the black hole seed mass). The \textit{central} AGN is assigned to the most massive black hole within a halo; all other AGNs within the same halo are tagged as \textit{satellite} AGNs.       

\subsection{Identification of AGN systems in MBII}
\label{AGN_identification}
We identify AGN systems based on a set of maximum distance criteria. In other words, we determine whether an AGN is a member of a system by defining a parameter denoted by $d_{\mathrm{max}}$. For a given system, $d_{\mathrm{max}}$ is defined as the maximum distance of every member AGN with respect to at least one other member AGN. 

We identify four types of AGN systems in the simulation as follows: 
%\tiziana{Recall you are actually using a 'real' FOF algorithm to identify haloes in the simulations - so you want to be careful how you mention FOF - it will be very confusing 
%to simulation people otherwise. I think here ou could simply state this is just a distance between AGNs correct? and there is no special physical value (e.g. linking length of 0.2 in FOF means virialized halo etc..- but here any distance is 'just a distance'. This is not a real comment - it is just a note about how language is used and given that you are using FOF haloes / simulations the language can be ambigous if you also refer to this as linking length or FOF}
\begin{itemize}
    \item \textit{Physical} AGN system: An AGN system where each member AGN is within a maximum comoving distance $d_{\mathrm{max}}$ w.r.t at least one other member AGN.% We investigate two different types of AGN systems defined as the following:
    
    \item \textit{Redshift-space} AGN system: This is similar to a \textit{Physical} AGN system except that the line of sight coordinate is in redshift space~(distorted due to peculiar velocities of the member AGNs).  
    
    \item \textit{Projected} AGN system: An AGN system where each member AGN is within a maximum comoving distance $d_{\mathrm{max}}^{\parallel}$ parallel to the plane of the sky w.r.t at least one other member AGN. In addition, the system must have a maximum velocity separation~(in this work we consider velocity separations $<2000~\mathrm{km/s}$ and $<600~\mathrm{km/s}$) along the line-of-sight\footnote{We adopt maximum velocity separations  $v_{\mathrm{max}}=2000~\mathrm{km/s}$ and $v_{\mathrm{max}}=600~\mathrm{km/s}$ from \cite{2006AJ....131....1H} and \cite{2011ApJ...737..101L} respectively.}; in other words, the maximum redshift space separation of the line of sight coordinate must be $v_{\mathrm{max}}/a H(a)$ from at least one other AGN within the system. Here, $a$ is the scale factor and $H(a)$ is the Hubble parameter. To avoid confusion, we further emphasize that our definition is in contrast to the more general use of the term `projected', which is more typically used to describe systems of objects within small enough angular separations but may have arbitrarily large line of sight separations.
    
    \item \textit{Bound} AGN system: An AGN system where all the members are within the same halo~(FOF group), implying that they are gravitationally bound w.r.t each other.
    
    %\item \textit{Observable} AGN system: The detectibility of a \textit{Physical}/ \textit{Redshift-space} space AGN system in a spectroscopic survey is limited by the fibre-collision limit~().  AGN system where each 
    
\end{itemize}
\subsection{Multiplicity function}
\label{multiplicity_function_methods}
For a given AGN system, we define \textit{richness} $\mathrm{R}$ to be the number of member AGNs within the system. For a given maximum distance $d_{\mathrm{max}}$, the \textit{Multiplicity Function} is defined to be the volume density ($n^{\mathrm{system}}(\mathrm{R}$)) or surface density ($\Sigma^{\mathrm{system}}(\mathrm{R})$) of AGN systems per unit \textit{richness}. The surface density can be obtained by integrating the volume density over the redshift width using
\begin{eqnarray}
\Sigma^{\mathrm{system}}= \int_{z_i}^{z_f} n^{\mathrm{system}}(z) r^2 dz \frac{dr}{dz} sr^{-1}
\label{surface_from_volume}
\end{eqnarray}
where $r(z)$ is the comoving distance to redshift $z$, $z_i$~(initial redshift) to $z_f$~(final redshift) is the redshift interval over which the surface density is to be computed. For the most part, we are focusing on the redshift interval of $0.06<z<4$. This redshift range encompasses current and future surveys namely SDSS, BOSS, eBOSS, DESI.  %In this work, we consider redshift bins of width $0.5$.%\adam{~You could note whether there is anything special about 0.5, or if it is simply an arbitrary or unimportant choice.}  
\subsection{Halo Occupation Distribution~(HOD) modeling of the AGN Multiplicity function}
\label{HOD_sec}
The Halo Occupation Distribution~(HOD) is defined as the probability distribution $P(N|M_h)$ of the number of AGNs $N$ in a halo of mass $M_h$. $P(N|M_h)$ can be written as
\begin{equation}
P(N|M_h)=P_{\mathrm{cen}}(N_{\mathrm{cen}}|M_h)+P_{\mathrm{sat}}(N_{\mathrm{sat}}|M_h)
\end{equation}
where $P_{\mathrm{cen}}(N_{\mathrm{cen}}|M_h)$ and $P_{\mathrm{sat}}(N_{\mathrm{sat}}|M_h)$ are contributions from central and satellite AGNs respectively; $N=N_{\mathrm{cen}}+N_{\mathrm{sat}}$ where $N_{\mathrm{cen}}$ and $N_{\mathrm{sat}}$ are the occupation numbers of central and satellite AGNs respectively within a halo. The first order moments of the HODs are the mean occupation numbers of central and satellite AGNs defined as $\left< N_{\mathrm{cen}}\right>$ and $\left< N_{\mathrm{sat}}\right>$ respectively.
$P_{\mathrm{sat}}(N_{\mathrm{sat}}|M_h)$ is assumed to be a Poisson distribution. Note that it has been previously found~\citep{2018MNRAS.480.3177B} that for low occupations~($\left< N_{\mathrm{sat}}\right> < 1$), the satellite HODs can become narrower than Poisson i.e. $\left< N_{\mathrm{sat}}(N_{\mathrm{sat}}-1)\right> < \left< N_{\mathrm{sat}}\right>$. In such a scenario, our HOD modeling can, at best, only provide an \textit{upper limit} for the abundance of AGN systems.

Under the assumptions listed in the previous paragraph, $P(N|M_h)$ can be completely determined from the first order moments $\left< N_{\mathrm{cen}}\right>$ and $\left< N_{\mathrm{sat}}\right>$. In our previous paper~\citep{2019MNRAS.485.2026B}, we constrained $\left< N_{\mathrm{cen}}\right>$ and $\left< N_{\mathrm{sat}}\right>$ using the small scale clustering measurements of quasar pairs~\citep{2017MNRAS.468...77E}. A brief description of the parametrization as well as the final constraints on $\left< N_{\mathrm{cen}}\right>$ and $\left< N_{\mathrm{sat}}\right>$, are presented in Appendix \ref{mean_occupations_sec}; readers interested in more details are encouraged to refer to \cite{2019MNRAS.485.2026B}. In Sections \ref{eboss_HOD} and \ref{volume_limit_sec} , we shall use these first order moments to make HOD model predictions for multiplicity functions using the methodology described in the next paragraph.

From $P(N|M_h)$, we can determine the fraction $f_{>R}(M_h)$ of haloes hosting \textit{bound} AGN systems with $\textit{richness}\geq R$. For a sample of haloes with mass $M_h$ to $M_h+dM_h$, $f_{>R}(M_h)$ is given by 
\begin{equation}
    f_{>R}(M_h)=\int_{R}^{\infty}P(N=R|M_h)~dR
\end{equation}
The volume density of \textit{bound} AGN systems with \textit{richness} $>R$ is then given by
\begin{equation}
    n_g^{\mathrm{system}}(>R)=\int f_{>R}(M_h)~\frac{dn}{dM_h}~dM_h
    \label{HOD_model_eqn}
\end{equation}
where ${dn}/{dM_h}$ is the halo mass function adopted from \cite{2008ApJ...688..709T}. We can then use Eq.~(\ref{surface_from_volume}) to determine the surface density of \textit{bound} AGN systems.

%\aklant{In Section \ref{volume_limit_sec}, we shall use these first order } Readers interested in more details are encouraged to refer to \cite{2019MNRAS.485.2026B}. 

\section{AGN multiplicity functions in MBII}
\label{MF_sec}
In this section, our eventual objective is to quantify the abundances of AGN systems in MBII by presenting their multiplicity functions as defined in Section \ref{multiplicity_function_methods}. We begin by first looking at AGN pairs in MBII and comparing their abundances to existing constraints of SDSS pairs at $\sim100~\mathrm{kpc}$ to validate our simulation. We then proceed towards higher order AGN systems~(triples and quadruples in particular) at a wider range of scales, and identify a target scale for systems that are gravitationally bound. We then present the multiplicity function predictions and discuss possible impediments for their detections in upcoming surveys.     
 
\subsection{Abundance of \textit{projected} AGN pairs in MBII: Comparison with SDSS pairs}
\begin{figure}
\includegraphics[width=8cm]{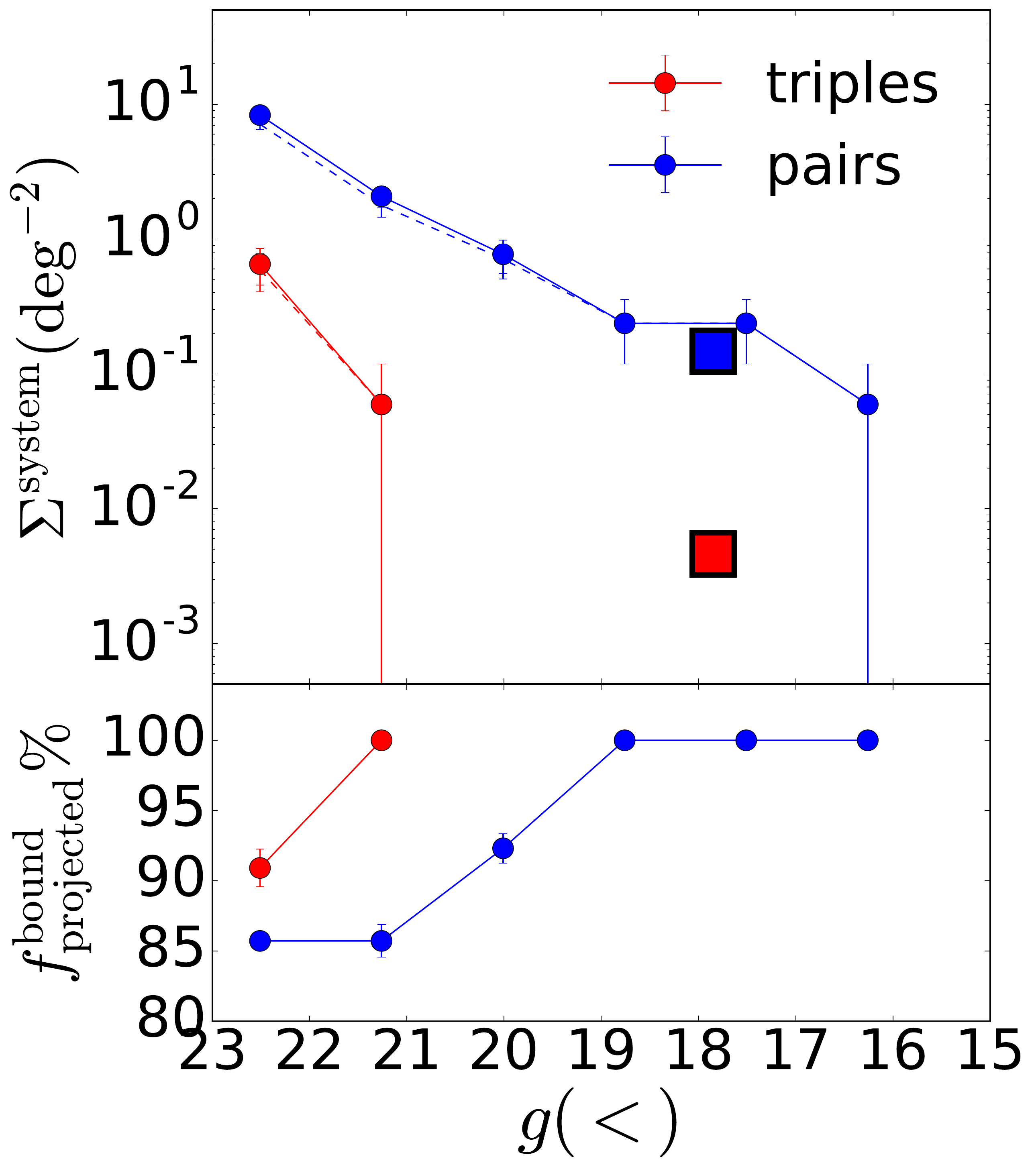}
\caption{\textbf{Top panels}: Filled circles connected by solid lines show the surface density of simulated AGN \text{projected} pairs~(blue) and triples~(red) over $0.02\lesssim z \lesssim0.15$ with $d^{\parallel}_{\mathrm{max}}=0.1~\mathrm{cMpc}/h$ and line of sight velocity difference of $600~\mathrm{km/sec}$. The dashed lines represent the surface densities of the subset of pairs~(blue) and triples~(red) satisfying the same foregoing criteria, but are also \textit{bound}. The filled squares represent observed surface densities of pairs~(blue) and triples~(red) inferred from \protect\cite{2011ApJ...737..101L}. \textbf{Bottom panels}: Percentage of the \textit{projected} pairs and triples~(plotted in top panel) that are \textit{bound}. The error-bars correspond to Poisson errors of the number counts.}
\label{MF_scale_richness_Liu}
\end{figure}
Before focusing on AGN triples and quadruples, we first look at abundances of projected AGN pairs at $\sim 100~\mathrm{ckpc}/h$ scales~(the prefix '$c$' denotes that the distance is in \textit{comoving} units), and compare them to constraints from SDSS~\citep{2011ApJ...737..101L}. In Figure \ref{MF_scale_richness_Liu}, we show surface densities of \textit{projected} AGN pairs and triples at $0.02\lesssim z \lesssim0.15$, with selection criteria similar to that of \cite{2011ApJ...737..101L} i.e. line of sight velocity differences of $600~\mathrm{kms/sec}$ and separations $\lesssim0.1~\mathrm{Mpc}/h$ on the plane of sky. We obtain these surface densities by computing the volume densities~(at each available snapshot between $z\sim0.02$ and $z\sim0.15$) and integrating them over $0.02\lesssim z \lesssim 0.15$ using Eq.~\ref{surface_from_volume}. Filled circles in blue show abundances of pairs in MBII; these are compared to surface densities inferred from observed AGN pairs~(blue square) in \cite{2011ApJ...737..101L} derived from samples of obscured AGNs in SDSS-DR7~\citep{2009ApJS..182..543A}. We find that the abundances of \textit{projected} pairs in MBII are broadly consistent with the observed AGN \textit{projected} pairs. This agreement is despite the fact that the observed samples contain only obscured AGNs whereas the simulated sample includes both obscured and unobscured AGNs, however obscured AGNs have also been found to dominate the overall AGN population~\citep{2015ApJ...802...89B}. We also show the simulated and observed triples as red circles and squares respectively, but abundances of AGN triples with $g<21$ are too small to be probed by the simulation given its volume. The bottom panels in Figure \ref{MF_scale_richness_Liu} show that $\gtrsim~85\%$ percent of these \textit{projected} pairs are \textit{bound} at $g<23$. The percentage increases with increasing brightness and for $g\lesssim19$, all \textit{projected} pairs are \textit{bound}; this implies that according to MBII, almost all \textit{projected} pairs in \cite{2011ApJ...737..101L} are expected to be \textit{bound}. Furthermore, the percentages of bound pairs are significantly higher compared to those derived using criteria from \cite{2006AJ....131....1H}~(shown later in Figure \ref{projected_bound_percentage}). This is simply due to significantly smaller plane of sky separations and velocity differences adopted by \cite{2011ApJ...737..101L} compared to \cite{2006AJ....131....1H}.

We have therefore established that the lowest order AGN systems~(pairs) in MBII have abundances that are comparable to existing observations. This serves as a validation for our tool~(MBII), and motivates us to proceed further and use it to look at the higher order AGN systems~(triples/ quadruples).
\subsection{At what scales are AGN systems gravitationally bound?}
%\subsection{Dependence on scale}
\begin{figure*}
\includegraphics[width=\textwidth]{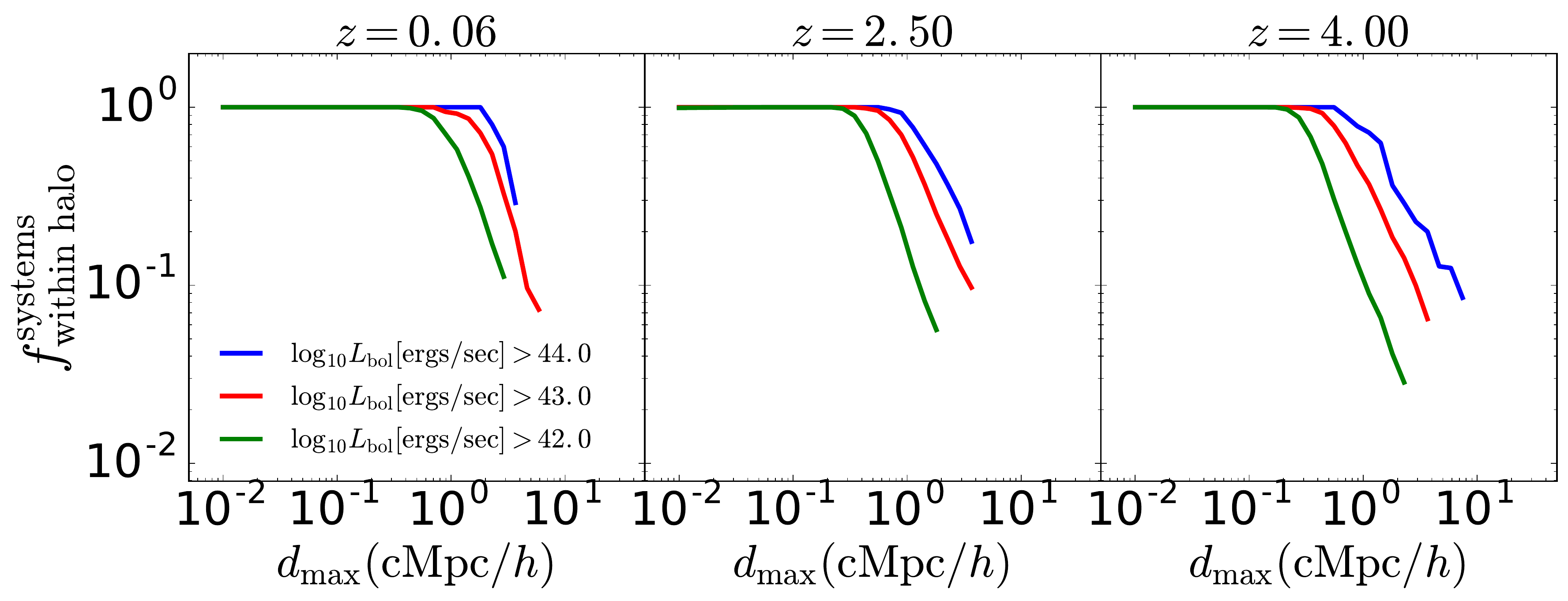}
\caption{$f^{\mathrm{system}}_{\mathrm{within~halo}}$ is the fraction of \textit{physical} AGN systems ($\texttt{RICHNESS}>1$) which are completely embedded within a common host halo, as a function of scale $d_{\mathrm{max}}$ defined as the maximum distance of separation between member AGNs.}
\label{MF_scale_fig}
\includegraphics[width=\textwidth]{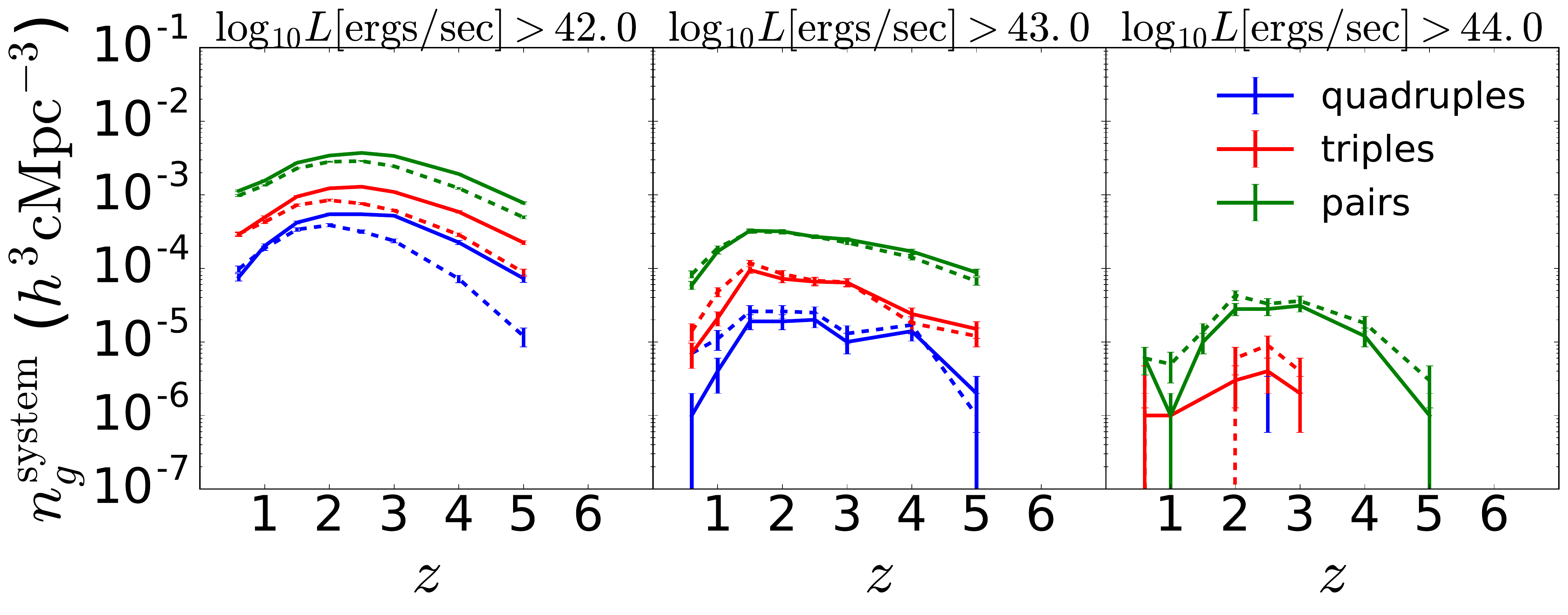}
\caption{Solid lines show the volume density of \textit{physical} AGN systems (pairs, triples and quadruples at $d_{\mathrm{max}}=0.7~\mathrm{cMpc}/h$) as a function of redshift from $z=0.6$ to $z=5$. The dashed lines show the volume density of \textit{bound} AGN systems~(i.e. within the same FOF halo) over the desired redshift range. The errorbars correspond to Poisson errors in the number counts.}
\label{MF_redshift}
\end{figure*}

%\tiziana{Fig1: Here you are using 4 thresholds in luminosity but in the rest of the paper you only have 3 - can you stick to 3 here too? (in any case the last one has nothing on (or you could add this set for > 45 to the > 44 plot if you have a strong reason to show it. Also, is there a real need to show all these redshifts? the plots look 'kind of the same' throughout. How about using just z=1 and 2 or actually (as you discuss far more redshifts later just one illustrative z (like z=2)?. This plot is far too large and busy for what it needs to convey and what the rest of the paper re-discusses anyway - the points you make from this plot in the text are (as you also discuss) pretty obvious/straightforward. 
%More minor but would be good to do: reset axis ranges - both x and y - so the are way narrower - only got out to 10 cMpc and start at 0.01 - also the y axis can you rest it to max 0.01 - there is too much white space in the plots and the curves are all squeezed / on top of each other. Please note 'true' is no really appropriate - use 'physical' or similar instead (i changed it the first time in the text but should be done throughout. Also here instead of labels for singlets, pairs and triples etc.. you should make smaller labers and call it R=1,2,3,4 etc.. or just mention it in the caption.}
We now look at AGN systems~(pairs, triples and beyond) for a wider range of scales~($\sim10~\mathrm{ckpc}/h-10~\mathrm{cMpc}/h$) quantified by $d_{\mathrm{max}}$~(see Section \ref{AGN_identification}). Our main objective is to prescribe a scale for targeting those \textit{physical} AGN systems which are gravitationally \textit{bound}~(i.e. all members belong to the same FOF halo). Figure \ref{MF_scale_fig} shows the fraction ($f^{\mathrm{systems}}_{\mathrm{within~halo}}$) of \textit{physical} AGN systems~($R>1$) completely embedded within a single host halo, as a function of scale $d_{\mathrm{max}}$~(note that we are considering \textit{physical} AGN systems at length scales $\gtrsim10~\mathrm{ckpc}/h$, which is roughly 10 times larger than the spatial resolution/ gravitational smoothing length of the simulation. We do this to minimize the possibility of our predictions to be affected by the finite spatial resolution of the simulation). Different panels show different redshifts ranging from $z\sim0$ to $z\sim4$ and different colors correspond to different luminosity cuts.  At small enough scales, all the \textit{physical} AGN systems are \textit{bound}~($f^{\mathrm{systems}}_{\mathrm{within~halo}}\rightarrow1$). Beyond a certain scale~(which we will refer to as $d_{\mathrm{max}}^{0}$), \textit{physical} AGN systems start to include members across different halos leading to a sharp drop in $f^{\mathrm{systems}}_{\mathrm{within~halo}}$ for $d_{\mathrm{max}}>d_{\mathrm{max}}^{0}$.

$d_{\mathrm{max}}^{0}$ depends on the typical sizes of haloes which host AGNs. The trends in Figure \ref{MF_scale_fig} show that $d_{\mathrm{max}}^{0}$ increases with luminosity at fixed redshift because more luminous AGNs live in more massive haloes~\citep{2017MNRAS.466.3331D,2019MNRAS.485.2026B}. Likewise, $d_{\mathrm{max}}^{0}$ decreases with increasing redshift~(at fixed luminosity) because host haloes become smaller at higher redshift. Overall, $d_{\mathrm{max}}^{0}$ ranges from $\sim0.5-2~\mathrm{cMpc}/h$ depending on the luminosity and redshift. 
%\adam{~Did we define ckpc anywhere? Perhaps we should define it at the end of the Introduction?}. 

The foregoing motivates us to focus on $d_{\mathrm{max}}\sim0.5-2~\mathrm{cMpc}/h$. For instance, in Figure \ref{MF_redshift} we select a value within this range, specifically $d_{\mathrm{max}}\sim0.7~\mathrm{cMpc}/h$ to show the volume density of AGN systems. The solid lines show the volume density of \textit{physical} AGN systems i.e. pairs, triples and quadruples~($R=2,3,4$) as a function of redshift. These can be compared to the dashed lines which represent \textit{bound} AGN systems. The volume density of \textit{physical} and \textit{bound} systems are similar for $L_{\mathrm{bol}}>10^{43,44}~\mathrm{ergs/sec}$ AGNs. We therefore hereafter~(unless stated otherwise) present results for $d_{\mathrm{max}}=0.7~\mathrm{cMpc}/h$; with this choice, we are primarily targeting \textit{physical} AGN systems which are within the same halo~(and are therefore bound). Note that our choice of scale is also comparable to the scales of observed AGN systems \citep{2007ApJ...662L...1D,2008ApJ...678..635M,2013MNRAS.431.1019F,2015Sci...348..779H}.

We also see in Figure \ref{MF_redshift}  that AGN systems are most abundant at the epoch $1.5\lesssim z \lesssim3$. In this epoch, the volume density of AGN triples with $L
_{\mathrm{bol}}>10^{44}~\mathrm{ergs/sec}$~(rightmost panel) is $\sim10^{-6}~h^3\mathrm{cMpc}^{-3}$.
\subsection{Dependence of AGN multiplicity function on $R$}
\label{MF_dependence_on_R_sec}
\begin{figure*}
\includegraphics[width=\textwidth]{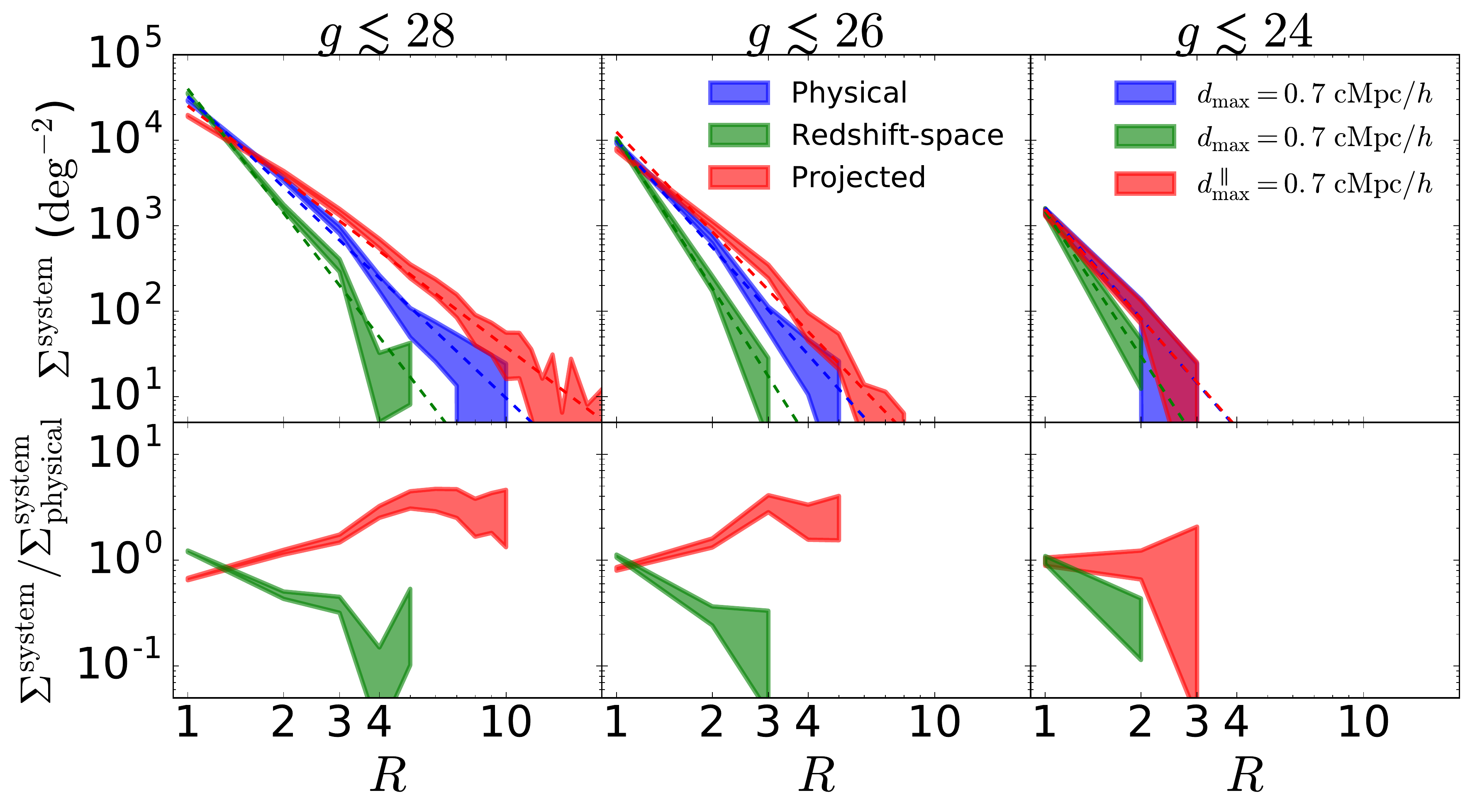}
\caption{\textbf{AGN multiplicity function}:- \textbf{Top panels:} $\Sigma^{\mathrm{system}}$ is the surface density of \textit{physical}, \textit{redshift space} and \textit{projected}~($v_{\mathrm{max}}=2000~\mathrm{km/sec}$) AGN systems between redshifts $0.06\lesssim z\lesssim4$ plotted as a function of \textit{richness} $\mathrm{R}$ (number of AGNs in a system) for different flux limits in the g-band. The middle and right panels roughly correspond to the expected flux limits of the LSST~\protect\citep{2009arXiv0912.0201L} and DESI~\protect\citep{2016arXiv161100036D} surveys respectively. Additionally, we also show~(left panel) AGN systems 2 magnitudes fainter than the LSST limit at which MBII contains \textit{physical} AGN systems with upto $\sim10$ members. The dashed lines are best fit power-laws~(coefficients are provided in Table \ref{power_law_fits_table}). \textbf{Bottom panels:} Ratio of surface density of \textit{projected} and \textit{redshift-space} systems with respect to \textit{physical} AGN systems as a function of $R$. The scatter corresponds to Poisson error.}
\label{MF_scale_richness}
\end{figure*}

\begin{table}
%\caption{Parameters for fitting results of \texttt{SAMPLE-TREE} under $M_h>10^{11}M_{\odot}h^{-1}$}
\centering
\begin{tabular}{|c|c|c|}
$g(<)$&$\alpha$ & $\Sigma_1$\\
\hline
& Physical &\\
\hline
28 & $-3.5$ & $3.2\times10^4$\\
26 & $-4.1$ & $1.0\times10^4$\\
24 & $-4.2$ & $1.6\times10^3$\\
\hline
& Redshift-space &\\
\hline
28 & $-4.8$ & $4.0\times10^4$\\
26 & $-5.8$ & $1.1\times10^4$\\
24 & $-5.6$ & $1.4\times10^3$\\
\hline
& Projected &\\
\hline
28 & $-2.8$ & $2.5\times10^4$\\
26 & $-3.9$ & $1.2\times10^4$\\
24 & $-4.2$ & $1.5\times10^3$\\
\hline
\end{tabular}

\caption{Best fit values of power-law coefficients $\alpha$ and $\Sigma_1$ for various types of AGN systems. $g(<)$ is the threshold magnitude in the $g$ band. The power law fits are shown as dashed lines in Figure \ref{MF_scale_richness}.}
\label{power_law_fits_table}
\end{table}

\begin{figure*}
\includegraphics[width=\textwidth]{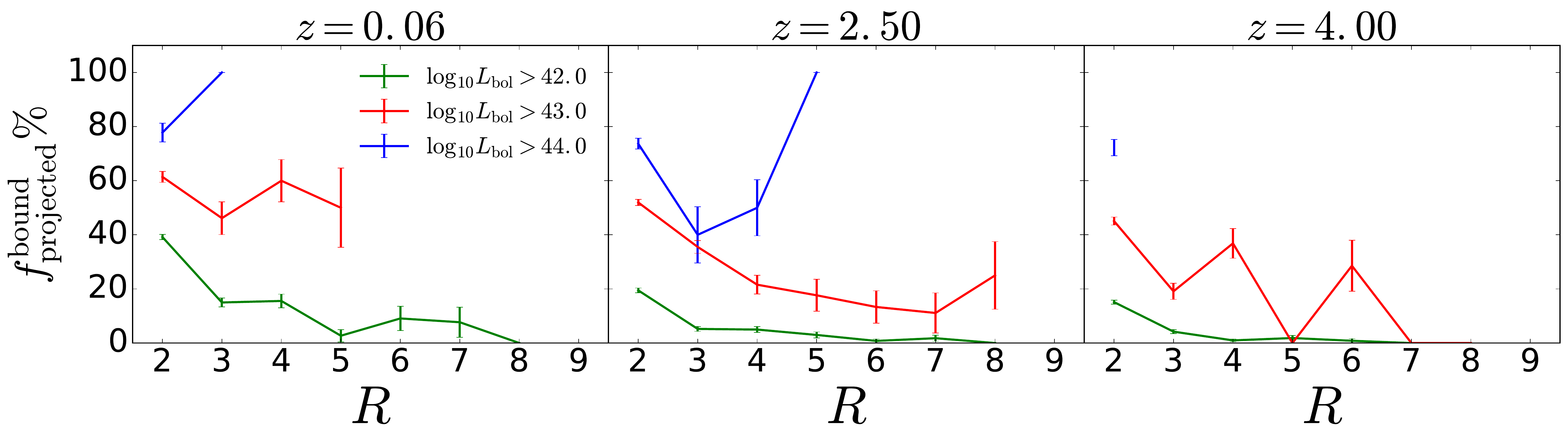}
\caption{The percentage of the \textit{projected}~($d^{\parallel}_{\mathrm{max}}=0.7~\mathrm{cMpc}/h$ and $v_{\mathrm{max}}=2000~\mathrm{km/sec}$) AGN systems that are \textit{bound}~(all members belong to the same host halo) versus richness. The errorbars correspond to Poisson errors in the number counts of \textit{projected} systems and \textit{bound} systems  .}%\tiziana{these HUGE/blocky colored areas look very weird/ugly. Should you use just points with error bars? also squeeze the y axis v. significantly so the plot is overall long and thin?}}
\label{projected_bound_percentage}
\end{figure*}

%\tiziana{you don't say what the left panel is - why you picked that magnitude. Also this caption is very long and a lot of the info is on the plot. Please try and shorten it. This applies to many other captions too. They repeat a ton of info that is already in the text and most of all on the plot itself}. 

With $\sim0.7~\mathrm{cMpc}/h$ identified as the target scale for gravitational bound AGN systems, we finally present the full multiplicity functions of MBII AGNs as a function of richness $R$. Figure \ref{MF_scale_richness} shows the surface density of AGN systems as a function of $R$. The g-band apparent magnitude limits have been chosen to be consistent with the expected detection limits of upcoming surveys such as DESI for $g<24$~\citep{2016arXiv161100036D}, LSST~\citep{2009arXiv0912.0201L} for $g<26$, and HST~\citep{2012OptEn..51a1011L} for $g<28$. $g<24$ is also representative of surveys such as  JPASS~\citep{2014arXiv1403.5237B}, SPHEREX~\citep{2014arXiv1412.4872D} and PAU~\citep{2019MNRAS.484.4200E}. Note that observationally, AGN systems are typically confirmed via spectroscopic follow-up, and spectroscopic surveys such as eBOSS \citep{2016AJ....151...44D}, DESI~(and possible future follow-up to LSST) will be limited in their detectibility due to the fibre-collision limit. We shall discuss the effect of these limitations in Section \ref{fibre_collision_sec}. In this section, we will focus on the \textit{intrinsic} abundance of AGN systems at various depths encompassing the imaging areas of HST, DESI \citep[provided by the Legacy Surveys; see, e.g.][]{2018arXiv180408657D} and the LSST.  

The surface densities in Figure \ref{MF_scale_richness} are calculated by integrating the volume density over redshifts ranging from $0.06\lesssim z\lesssim4$ using Eq.~(\ref{surface_from_volume}). Blue, green and red shaded regions correspond to \textit{physical}, \textit{redshift-space} and \textit{projected} AGN systems as defined in Section \ref{AGN_identification}. For the \textit{physical} and \textit{redshift-space} systems we use $d_{\mathrm{max}}=0.7~\mathrm{cMpc}/h$~(representative of \textit{bound} AGN systems as already discussed in Figure \ref{MF_redshift}). For the \textit{projected} systems, we use $d_{\mathrm{max}}^{\parallel}=0.7~\mathrm{cMpc}/h$~(maximum distance between members parallel to the plane of the sky). We find that the surface density of AGN systems exhibits a power-law decrease w.r.t increase in $R$. The power law may be described as
\begin{equation}
\Sigma^{\mathrm{system}}=\Sigma_1R^\alpha
\end{equation} 
where $\alpha$ is the power-law slope and $\Sigma_1$ is a coefficient measuring the number of isolated AGNs~(i.e. $R=1$). The power-law fits are shown as dashed lines, and the values of $\alpha$ and $\Sigma_1$ are tabulated in Table \ref{power_law_fits_table}. The exponents range from $\sim-3$ to $\sim-6$ depending on the type of system and the magnitude threshold. For instance, for projected AGN systems with $g<24,26$ we see that $\alpha \sim -4$, which implies that roughly $\sim10\%$ of AGNs are part of AGN systems~(pair or higher) with $d_{\mathrm{max}}^{\parallel}=0.7~\mathrm{cMpc}/h$. It is also worthwhile to note that if we reduce $d_{\mathrm{max}}^{\parallel}$ to $\sim30~\mathrm{ckpc}/h$, only $\sim1\%$ of AGNs are part of AGN systems~(pair or higher), and this is consistent with predictions from \texttt{EAGLE} hydrodynamic simulation~\citep{2019MNRAS.483.2712R}.

From the multiplicity functions, we can read off the abundances of triples/ quadruples at various depths. Let us first focus on the abundances of \textit{physical} triples/ quadruples. At $g<24$~(depth of DESI imaging), surface density of available \textit{physical} triples is $\sim10$ per $\mathrm{deg}^2$. Likewise, at $g<26$~(depth of LSST imaging) MBII predicts $\sim100$ \textit{physical} triples/ quadruples per $\mathrm{deg}^2$. Lastly, at $g<28$~(depth of HST imaging), MBII predicts $\sim700$ \textit{physical} triples/ quadruples per $\mathrm{deg}^2$.
%\textit{Physical} vs. \textit{redshift-space} AGN systems

The comparison between the multiplicity functions for \textit{physical} and \textit{redshift-space} AGN systems is of particular interest because observations can only access the latter. We find that multiplicity functions for \textit{redshift-space} AGN systems are significantly more suppressed~(by factors up to $\sim10$) compared to \textit{physical} AGN systems. This occurs because of the dispersion in peculiar velocities of AGNs at small~(one-halo) scales which leads to higher line of sight separations between member AGNs in redshift space as compared to real space. This implies that the sample of AGN systems obtained from a spectroscopic follow-up of \textit{projected} AGNs may not include all possible \textit{physical/bound} AGN systems. This can be corrected by simply choosing a larger line of sight~(redshift space) separation $d_{\mathrm{max}}^{\perp}$ compared to $d_{\mathrm{max}}^{\parallel}$. We compare the multiplicity functions for \textit{physical} and \textit{redshift-space} systems for a wide range of ratios $d_{\mathrm{max}}^{\perp}/d_{\mathrm{max}}^{\parallel}$; we find that in order to match the multiplicity functions for \textit{redshift-space} and \textit{physical} AGN systems, we would require $d_{\mathrm{max}}^{\perp}/d_{\mathrm{max}}^{\parallel}\sim 5$. In other words, for $d_{\mathrm{max}}^{\perp}/d_{\mathrm{max}}^{\parallel}> 5$ and $d_{\mathrm{max}}^{\parallel}\sim0.7~\mathrm{Mpc}/h$, the chosen line of sight separation is too large causing the multiplicity functions to be contaminated by systems that are unbound. On the other hand, for $d_{\mathrm{max}}^{\perp}/d_{\mathrm{max}}^{\parallel}<5$ and $d_{\mathrm{max}}^{\parallel}\sim0.7~\mathrm{Mpc}/h$, the chosen line of sight separation is too small, leading to significant number of bound systems that would be missed. 

\subsubsection{How to target gravitationally bound AGN systems in observations?}

We now have all the information required to target \textit{bound} AGN systems in observations. To ensure that the members are indeed bound~(i.e. belonging to the same halo), the comoving separation should be $\lesssim 0.7~\mathrm{cMpc}/h$. This corresponds to maximum angular separations ranging from  $\lesssim 100~\mathrm{arcsec}$ at $z\sim0.6$ to $\lesssim30~\mathrm{arcsec}$ at $z\sim4$. Along the line of sight co-ordinate, $\lesssim 0.7~\mathrm{cMpc}/h$ corresponds to maximum spectroscopic redshift separations ranging from $\delta z\lesssim5\times10^{-4}$) at $z\sim0.6$ to $\delta z\lesssim1\times10^{-3}$ at $z\sim4$. However, to compensate for the broadening of the line of sight separations due to peculiar velocity dispersions, one must increase the line of sight separation by factors $\sim5$, implying maximum spectroscopic redshift separations ranging from $\delta z\sim1.5\times10^{-3}$ at $z\sim0.6$ to $\delta z\sim7\times10^{-3}$ at $z\sim4$. The corresponding values of the maximum line of sight velocity difference range from $\lesssim150~\mathrm{km/sec}$ at $z\sim0.6$ to $\lesssim200~\mathrm{km/sec}$ at $z\sim4$. By applying the foregoing conditions, we can ensure that a spectroscopically confirmed sample of observed AGN systems shall correspond to a complete sample of \textit{bound} AGN systems.  

The inferred velocity separations for \textit{bound} AGN systems is about 10 times smaller than the $2000~\mathrm{km/sec}$. This limit was adopted by \cite{2006AJ....131....1H}, which includes velocity differences up to $\sim500~\mathrm{km/sec}$ expected from peculiar velocities, as well as the redshift uncertainties caused by blue-shifted broad lines~\citep{2002AJ....124....1R} of the member AGNs. Therefore, not all \textit{projected} systems are \textit{bound}. We therefore naturally expect more \textit{projected} AGN systems compared to \textit{physical/ redshift-space} AGN systems, as clearly seen in Figure \ref{MF_scale_richness}. The solid lines in Figure \ref{projected_bound_percentage} show the percentage of \textit{projected}~($v_{\mathrm{max}}=2000~\mathrm{km/sec}$, $d_{\mathrm{max}}=0.7~\mathrm{cMpc}/h$) systems that are \textit{bound}. We see that for $L_{\mathrm{bol}}>10^{42}~\mathrm{ergs/sec}$~(solid blue line), $\sim40\%$ and $\sim20\%$ of AGN systems are \textit{bound} at $z=0.06$ and $z=2.5$ respectively. We also see that at higher luminosities, a higher percentage of \textit{projected} AGN systems are \textit{bound}~($\sim50\%$ for $L_{\mathrm{bol}}>10^{43}~\mathrm{ergs/sec}$). As a result, the difference between the multiplicity functions of \textit{physical} and \textit{projected} AGN systems decreases in Figure \ref{MF_scale_richness} with increasing brightness. This is expected because with increasing luminosity, AGNs becomes rarer, which results in a decreased likelihood of a chance superposition of otherwise unbound AGNs on a projected plane. 
\subsubsection{Predictions for $g<22$~(eBOSS depths) AGN systems using HOD modeling}
\label{eboss_HOD}
%\subsection{Implications for eBOSS-CORE}
\begin{figure}
\includegraphics[width=8cm]{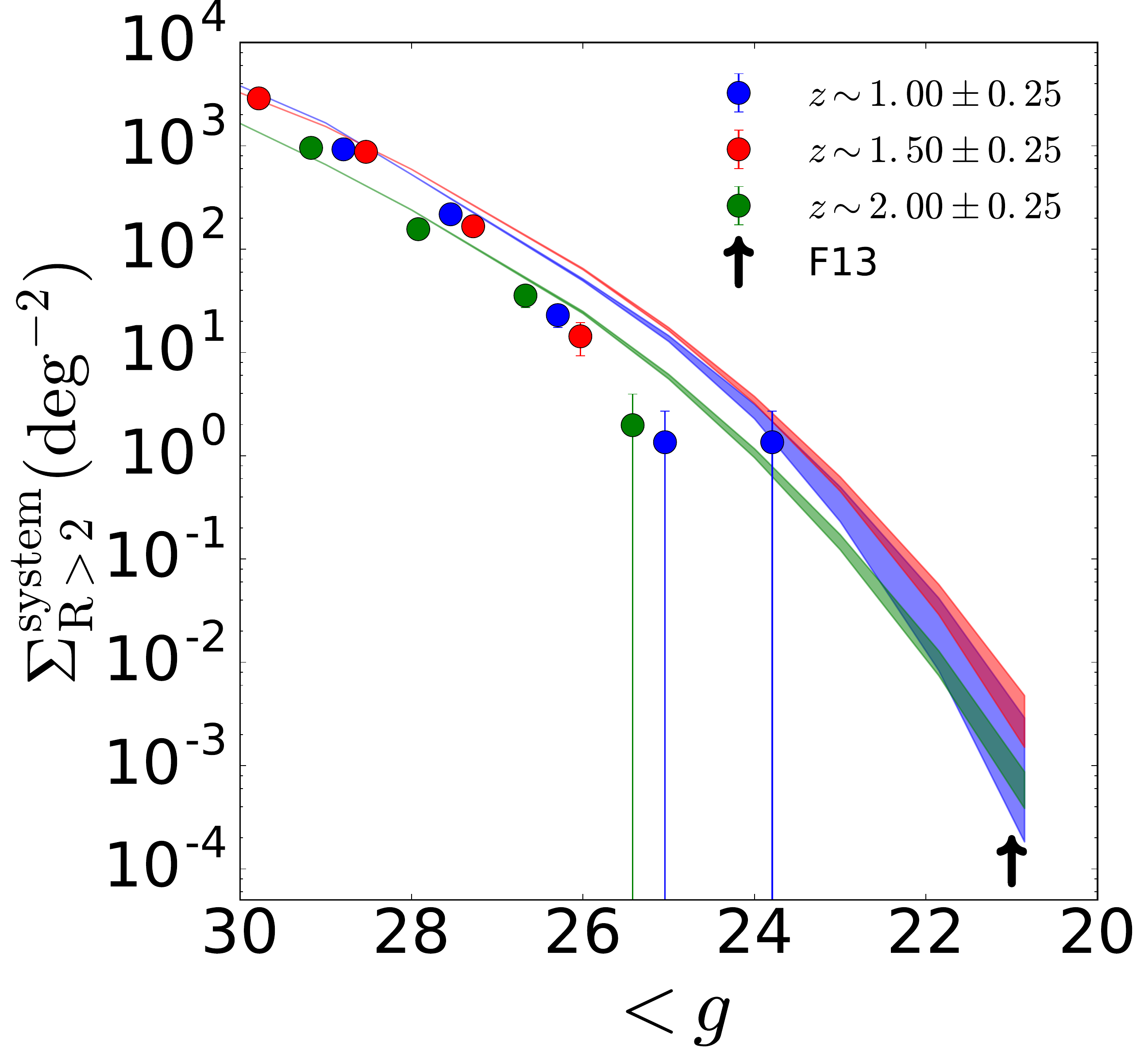}
\caption{Surface density of \textit{physical} AGN systems with $R>2$~(triples and higher-order systems) with $d_{\mathrm{max}}\sim0.7~\mathrm{cMpc}/h$~(angular separations $\lesssim50~\mathrm{arcsec}$) as a function of limiting $g$ band magnitude. The different colors correspond to different redshift bins within the target selection of eBOSS-CORE quasars, further limited to $0.9\lesssim z \lesssim2.2$ where optical selection of quasars is most efficient. The shaded regions correspond to the predictions~(upper limit) of a HOD model constrained by recent small-scale clustering measurements \protect\citep{2017MNRAS.468...77E,2019MNRAS.485.2026B}. The black arrow is a lower limit inferred from the observation of an AGN triple in \citet{2013MNRAS.431.1019F}.} 
\label{MF_scale_magnitude}
\end{figure}
MBII directly probes AGN triples/ quadruples up to $g\sim24$ over $0.06\lesssim z \lesssim 4$. In order to reach eBOSS like depths~($g\sim22$), we use the HOD model built in \cite{2019MNRAS.485.2026B}~(see Section \ref{HOD_sec} and Appendix \ref{mean_occupations_sec} for more details) and make predictions beyond the simulated regime. These are shown as shaded regions in Figure \ref{MF_scale_magnitude}\footnote{We use a Poisson distribution to model the HODs for satellite AGNs; in reality, the satellite occupation distribution may be narrower than Poisson at the rare luminous end, in which case the shaded regions in Figure \ref{MF_scale_magnitude} would correspond to upper limits for the surface density.}. The black arrow corresponds to the lower limit inferred from the current state of observations i.e. 1 triple found~\citep{2013MNRAS.431.1019F} so far during the follow-up search in the sample from SDSS-DR6 \citep{2009ApJS..180...67R}. Our predictions therefore do not conflict with observations so far. For $g<22$~(the approximate depth eBOSS-CORE), we predict $\sim10^{-2}$ systems per $\mathrm{deg}^{2}$, implying $\sim100$ AGN triples~(and higher-order systems) over the entire (imaging) area ($\sim7500~\mathrm{deg}^{2}$) of eBOSS.  

\subsection{Effect of volume limit on multiplicity functions}
\label{volume_limit_sec}
\begin{figure*}
\includegraphics[width=\textwidth]{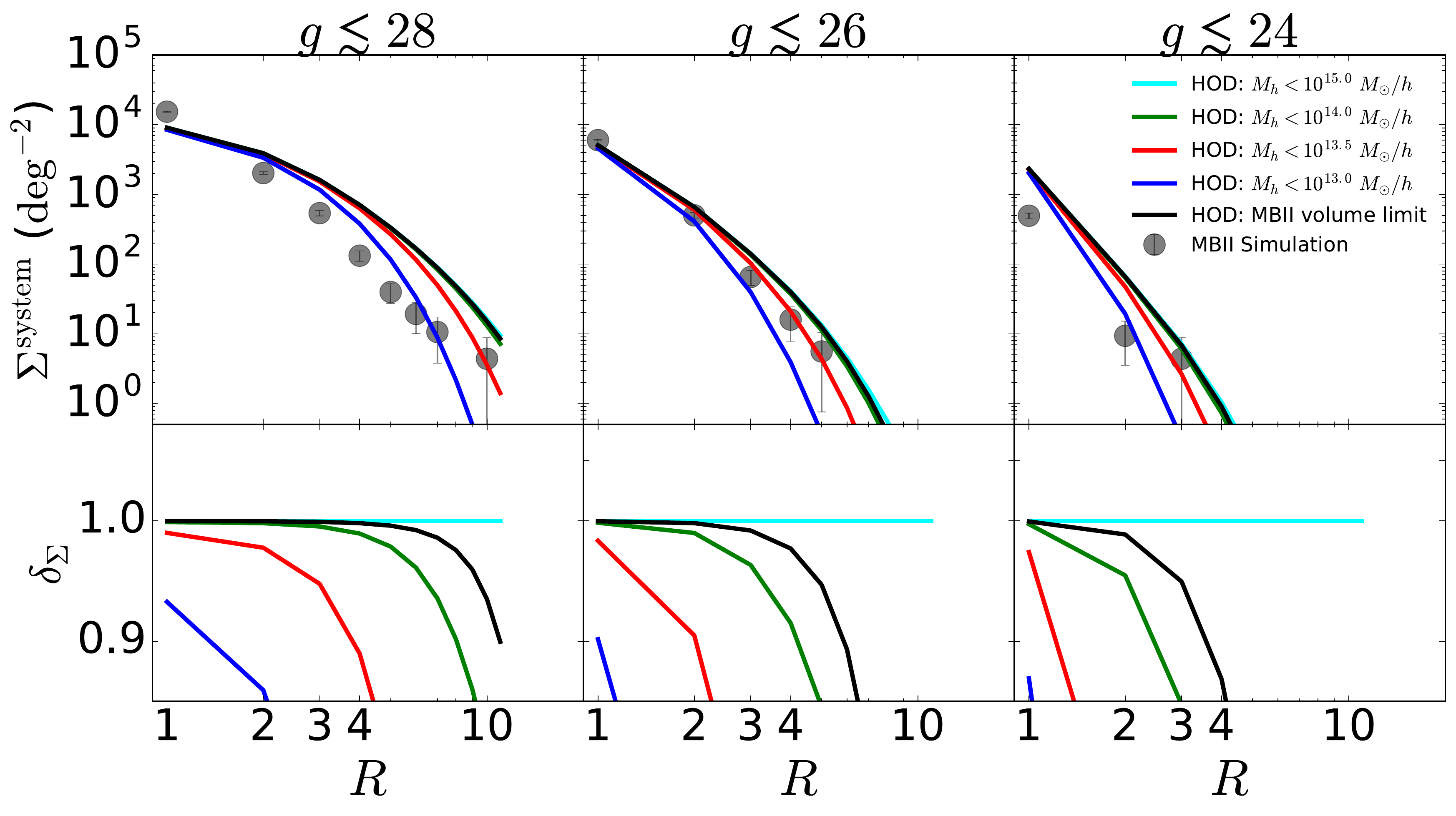}
\caption{\textbf{Top Panels}: Solid lines show the HOD model predictions for the surface densities of \textit{bound} AGN systems over $0.6\lesssim z\lesssim2$. Different colors correspond to various upper limits in the halo mass in Eq.~(\ref{HOD_model_eqn}), representative of different simulation volumes. Filled circles are the simulation predictions with errorbars showing Poisson errors. \textbf{Bottom Panels}: $\delta_{\Sigma}$ is the ratio between HOD model multiplicity functions with respect to that of $M_h<10^{15}~M_{\odot}/h$.}
\label{multiplicity_function_ref}
\end{figure*}       
The multiplicity functions in MBII can be potentially underestimated due to finite volume in the simulation, particularly for the AGN systems that are the brightest and have the highest richness. This is because the rarest massive haloes~($M_H\gtrsim10^{14}~M_{\odot}/h$) are too rare to be captured by the simulation. This is typical of all hydrodynamic simulations, and at its resolution MBII is amongst the largest set of cosmological hydrodynamic simulations run till date.

In order to estimate the possible suppression due to the finite volume, we use HOD modelling, wherein we effectively populate central and satellite AGNs on the halo mass function of \cite{2008ApJ...688..709T}, which has been calibrated to $~5\%$ accuracy with N-body simulations over  $10^{11}\lesssim M_h\lesssim10^{15}~M_{\odot}/h$. As described in Section \ref{HOD_sec}, HOD model was built using small scale clustering constraints at $0.6\lesssim z \lesssim 2$; we therefore select this redshift range to provide our estimate. We show in Figure \ref{multiplicity_function_ref}: top panels HOD model predictions of the multiplicity functions~(surface density) of \textit{bound} AGN systems over $0.6\lesssim z \lesssim 2$. We have modelled the effect of finite volume as an upper limit of halo mass~($M_h(<)$) in the integral of Eq.~(3). Solid lines~(blue, green, red and cyan) show predictions for various possible volumes corresponding to $M_h(<)=10^{13,13.5,14,15}~M_{\odot}/h$. As expected, the differences amongst the solid lines become more pronounced at higher richness and brighter magnitudes. We see that the multiplicity functions converge when the upper limit is $M_h\sim10^{15}~M_{\odot}/h$~(cyan line), with no significant contribution coming from higher mass ($M_h\gtrsim10^{15}~M_{\odot}/h$) haloes. 
We now seek an HOD model prediction that is representative of the MBII volume.
MBII (given its volume) can probe up to halo masses of $M_h\sim6\times 10^{13},1\times10^{14},2\times10^{14},4\times10^{14}~M_{\odot}/h$ at $z=2,1.5,1.0,0.6$ snapshots respectively. Accordingly, we choose these different upper limits at the respective redshifts to compute an HOD model prediction to the multiplicity function for the MBII volume over $0.6<z<2$. This is shown by the solid black line. Comparing the solid black line and the cyan line then provides an estimate of the suppression in the MBII multiplicity functions due to its finite volume. In the bottom panel of Figure \ref{multiplicity_function_ref} we show $\delta_{\Sigma}$, which is defined as the ratio of the HOD model multiplicity functions with respect to that of $M_h<10^{15}~M_{\odot}/h$. The black lines in the bottom panels correspond to the ratio between black and cyan lines of the top panel. They show that according to HOD modeling, the number of triples/ quadraples in MBII may be suppressed~(due to finite volume) by factors of $\sim0.85$, $0.97$, $\gtrsim0.99$ for $g<24,26,28$ respectively. Overall, this implies that the finite simulation volume of MBII leads to a very marginal suppression~($\lesssim15\%$) on the predictions of its multiplicity functions.

%We now put our results in the context of the ongoing eBOSS-CORE QSO~\citep{2015ApJS..221...27M} survey. Filled circles in  Figure \ref{MF_scale_magnitude} show the surface density of quasar systems with $R>2$~(triples and higher-order systems) as a function of apparent magnitude. The various redshits are a subset of the redshift range targeted by eBOSS~($0.9\lesssim z\lesssim2.2$). Given its volume, MBII can directly probe up to magnitudes of $g\sim25$, and predicts $\sim1$ system per $\mathrm{deg}^2$ for $g<25$. 

%These exponents are useful in making quick estimates of the expected number of quasar systems of a given \texttt{RICHNESS}, for a quasar sample of a given size. For e.g., the upcoming eBOSS sample will probe quasars with $L_{\mathrm{bol}}\gtrsim10^{45}~\mathrm{ergs/sec}$ ($g\lesssim22$). While these luminosities are larger than the range probed by MBII, we can certainly say that $\alpha\lesssim-4$ for eBOSS quasars (i.e. lower than $\alpha\sim-4$ reported for $L_{\mathrm{bol}}>10^{44} \mathrm{ergs/sec}$ MBII quasars). Upon completion, eBOSS is expected to comprise of $\sim10^{6}$ quasars; we can therefore estimate (assuming most quasars are singlets, i.e. $\Sigma_1\sim10^{6}$) the expected number of triples and quadruples in the eBOSS sample to be $~\lesssim 40,000$ and $\lesssim 1500$ respectively.    

\subsection{Effect of spectroscopic fibre collisions on the multiplicity functions}

\begin{figure*}
\includegraphics[width=\textwidth]{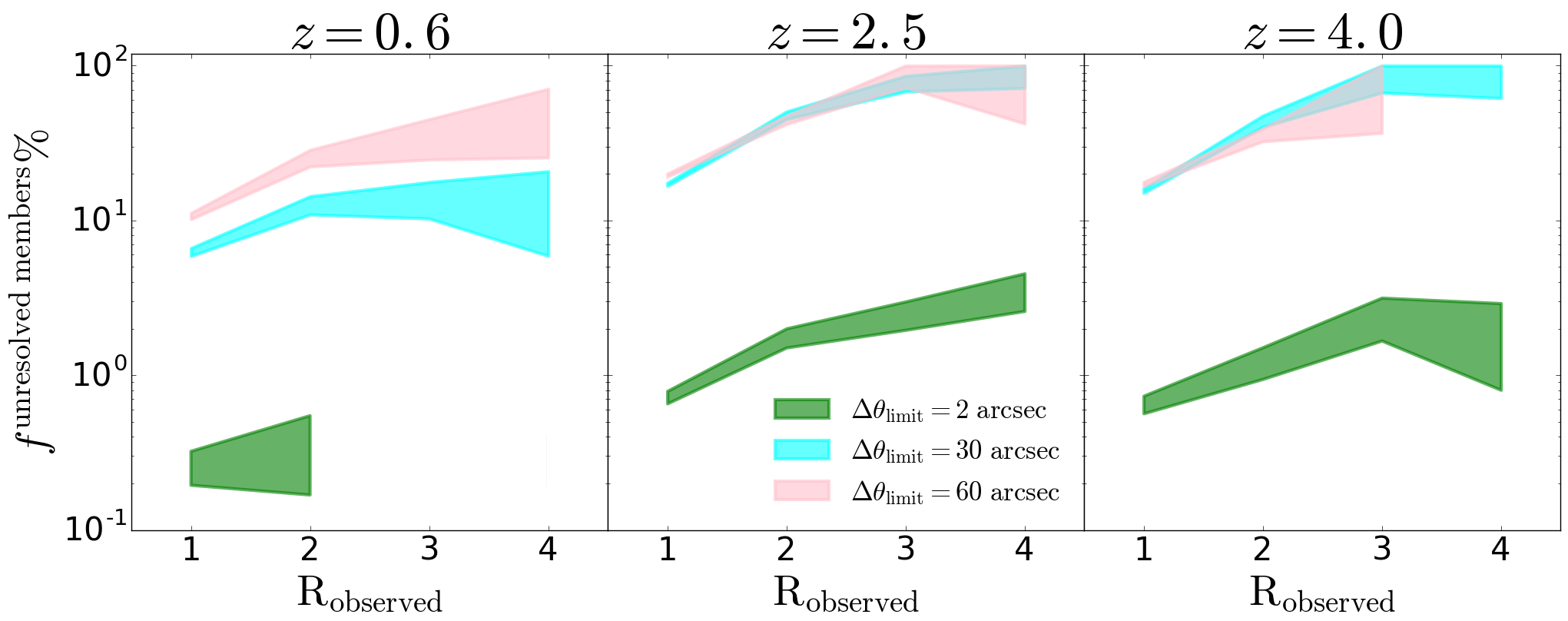}
\caption{$f^{\mathrm{unresolved~members}}$ is the percentage of AGN systems with \textit{richness} $R_{\mathrm{observed}}$ that have \textit{at least} one other extra member unresolved due to fibre collisions. $\Delta \theta_{\mathrm{limit}}$ is the minimum angular separation between two adjacent fibres. The shaded region shows the Poisson error. This plot shows AGN systems with $L_{\mathrm{bol}}>10^{42}~\mathrm{ergs/sec}$.}
\label{missing_members}
\end{figure*}

\begin{figure*}
\includegraphics[width=\textwidth]{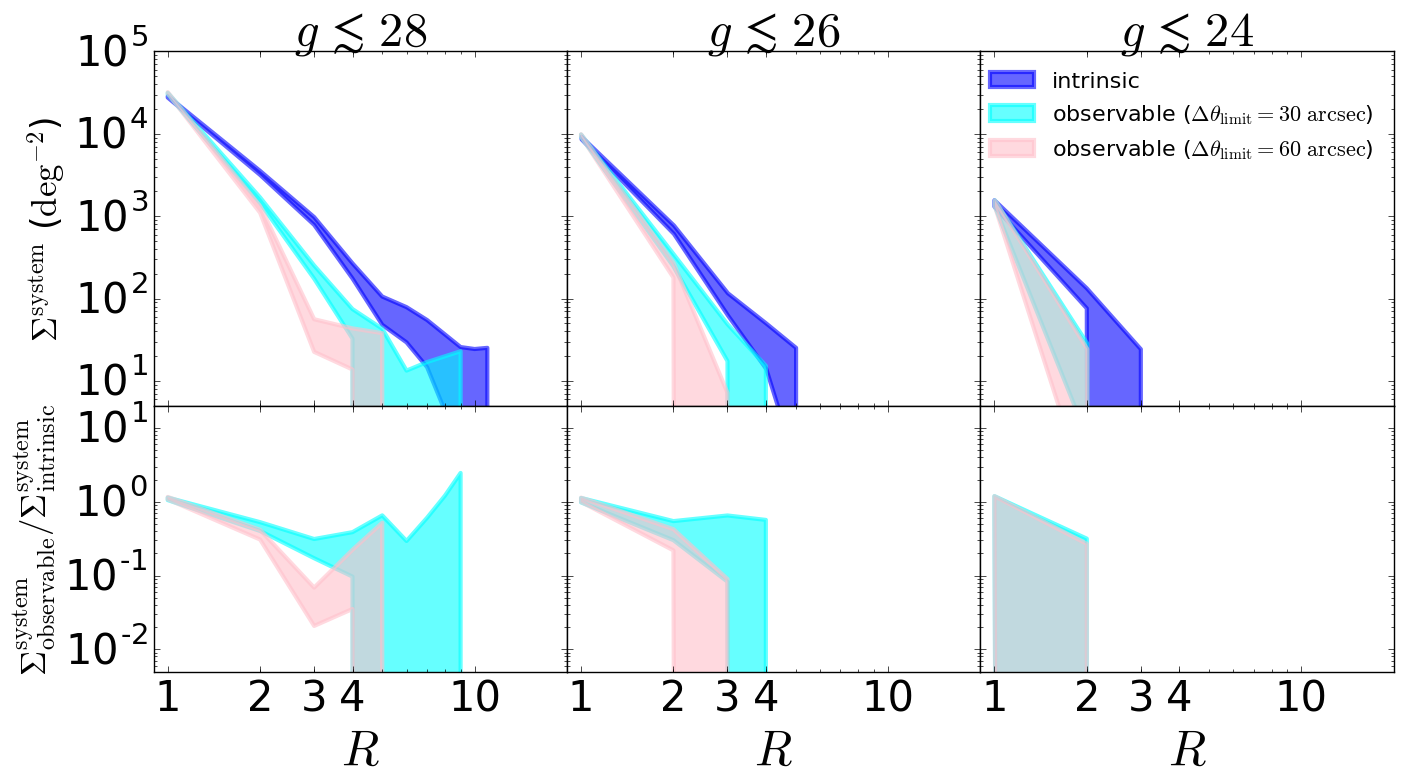}
\caption{\textbf{AGN multiplicity function with correction for fibre collisions}:- \textbf{Top panels:} $\Sigma^{\mathrm{system}}$ is the surface density of AGN systems~($d_{\mathrm{max}}=0.7~\mathrm{cMpc}/h$) over redshifts $0.06\lesssim z\lesssim4$, plotted as a function of \textit{richness} $\mathrm{R}$. The scatter corresponds to $1\sigma$ Poisson error. The blue region~($\Sigma^{\mathrm{system}}_{\mathrm{intrinsic}}$) corresponds to the \textit{intrinsic} abundance of \textit{physical} AGN systems as also shown in Figure \ref{MF_scale_richness}. The cyan and pink regions~($\Sigma^{\mathrm{system}}_{\mathrm{observable}}$) correspond to \textit{observed} abundances of AGN systems in the presence of a fibre-collision limit $\Delta \theta_{\mathrm{limit}}$~(see section \ref{fibre_collision_sec}). We assume a $30\%$ overlap between the areas of adjacent tiles. \textbf{Bottom panels:} Ratio between $\Sigma^{\mathrm{system}}_{\mathrm{observable}}$ and $\Sigma^{\mathrm{system}}_{\mathrm{intrinsic}}$.}
\label{MF_scale_richness_corrected}
\end{figure*}
\label{fibre_collision_sec}

As mentioned in Section \ref{MF_dependence_on_R_sec}, detectable AGN systems in spectroscopic surveys are subject to a minimum angular separation due to the \textit{fibre collision limit} i.e. the minimum distance between adjacent fibres on a single \textit{spectroscopic tile}~\citep[see, e.g.][for details on tiling algorithms]{2003AJ....125.2276B}. Here, we study its implications on the inferred multiplicity functions. We consider the following two crucial effects in our modeling.  
\begin{itemize}
\item \textbf{Loss of member AGNs due to spectroscopic fibre collisions}: To model this, we define a minimum angular separation, $\Delta\theta_{\mathrm{limit}}$~(hereafter referred as the \textit{fibre collision limit}), such that member AGNs at smaller separations cannot be distinguished from one another. We consider fibre collision limit of $60~\mathrm{arcsec}$, roughly representative of the SDSS, BOSS and eBOSS technical specifications~\citep{2003AJ....125.2276B,2013AJ....145...10D,2016AJ....151...44D}. Additionally, we also consider fibre collision limit of $30~\mathrm{arcsec}$, roughly consistent with a lower bound for DESI (this will however depend on exactly how the DESI focal plane is populated). 

\item \textbf{Recovery of member AGNs in regions with overlapping tiles:} We assume $30\%$ overlap between surface areas of neighboring \textit{tiles}, roughly representative of the layout of the tiles in the SDSS survey \citep{2003AJ....125.2276B}. A pair of AGNs in an area covered by overlapping tiles can be distinguished from one another despite having angular separations of less than $\Delta\theta_{\mathrm{limit}}$. Note, however, that our chosen $30\%$ overlap is a gross simplification for the DESI survey, for which the planned layout of tiles is quite complex, with multiple layers of spectroscopic tiles \citep[see Section 4 of][]{2016arXiv161100036D}.
\end{itemize}

Specifically, we consider each \textit{physical} AGN system with \textit{original richness} $R$ and $d_{\mathrm{max}}=0.7~\mathrm{cMpc}/h$ (shaded blue region in Figure \ref{MF_scale_richness}), calculate angular separations between their members, and reduce their richness by 1 unit for every distinct pair of members with angular separations below the resolution. We then randomly choose $30\%$ of the overlapping pairs and recover them i.e. for each of the recovered pair, we increase the richness of the parent system by 1 unit. We refer to the resulting richness of the system as $R_{\mathrm{observed}}$. $R_{\mathrm{observed}}$ is inevitably less than or equal to \textit{intrinsic} richness $R$.

Figure \ref{missing_members} shows the percentage ($f^{\mathrm{unresolved\ members}}~\%$) of AGN systems~($L_{\mathrm{bol}}>10^{42}~\mathrm{ergs/sec}$) with at least one extra unresolved member, for fibre collision limits of $30~\mathrm{arcsec}$~(cyan region) and $60~\mathrm{arcsec}$~(pink region). We see that about $\sim10\%$ of observed single AGNs at $z=0.6$~(leftmost panel) have unresolved members, which increases to $\sim20\%$ at $z=2.5,4$~(middle and rightmost panels); this is expected as the fibre collision limit~($\Delta\theta_{\mathrm{limit}}$) subtends a larger comoving separation at higher redshifts, leading to higher number of unresolved members. Additionally, we also see that higher the observed richness of a system, higher the probability of the system having additional unresolved members. For instance, at fibre collision limit of $30~\mathrm{arcsec}$, observed AGN triples and quadruples at $z=2.5$ and $z=4$ have $\gtrsim60\%$ chance of revealing additional unresolved members upon a possible future follow-up spectroscopic survey with lower fibre collision limit. In addition to the fibre collision limits of $30-60$ arcsec, it is also instructive to show $\Delta\theta_{\mathrm{limit}}=2~\mathrm{arcsec}$~(green region in Figure \ref{missing_members}). This is representative of angular resolution limits due to the point spread function~(PSF) of an observed source, and corresponds to $\lesssim70~\mathrm{ckpc}$ at $z<4$. We see that at $z=2.5,4$, about $\sim1-3\%$ of systems contain additional unresolved members; at $z=0.6$, only about $\sim0.3\%$ of systems contain additional unresolved members. Therefore the percentage~($f^{\mathrm{unresolved\ members}}~\%$) for the PSF limit of 2 arcsec is negligibly small compared to that of the fibre collision limits of $30-60~\mathrm{arcsec}$. Overall, this implies that the impact of the PSF limit on our inferred multiplicity functions for $d_{\mathrm{max}}=0.7~\mathrm{cMpc}/h$ is not very significant. In the next paragraph, we address the impact of the fibre collision limits on the multiplicity functions.

Figure \ref{MF_scale_richness_corrected} shows the impact of spectroscopic fibre collisions on the inferred multiplicity function. In the top panels, the blue region shows the \textit{intrinsic} multiplicity function~($\Sigma^{\mathrm{system}}_{\mathrm{intrinsic}}$) of \textit{physical} AGN systems~(as in Figure \ref{MF_scale_richness}); the cyan and pink regions show the multiplicity functions~($\Sigma^{\mathrm{system}}_{\mathrm{observable}}$) of the same set of AGN systems when corrected for fibre collision limits of $30~\mathrm{arcsec}$ and $60~\mathrm{arcsec}$ respectively. As expected, we see that with increasing $\Delta \theta_{\mathrm{limit}}$, we start to lose member AGNs and the abundance of higher order systems starts to get suppressed, leading to increasingly steep multiplicity functions. The bottom panels show the ratio between the observed multiplicity function and the intrinsic multiplicity function. We find that a fibre collision limit of $60~\mathrm{arcsec}$ suppresses the observed number of AGN triples/ quadruples by factors $\lesssim 0.2$); this implies that $\lesssim 20~\%$ of all available triples/quadruples are expected to be detected after spectroscopic follow-up. Note also that in regions with overlaps across multiple surveys such as SDSS and/or eBOSS and/or DESI, more missing members can be recovered. We do not model this effect as MBII does not directly probe AGN systems with magnitudes brighter than the SDSS and eBOSS detection limits (e.g., see Figure \ref{MF_scale_magnitude}).

\section{Origin and Environment of Quasar systems in MBII:}
\label{selected_systems_sec}

\begin{figure*}
\addtolength{\tabcolsep}{-5pt}
\begin{tabular}{cccc}
\includegraphics[width=5cm]{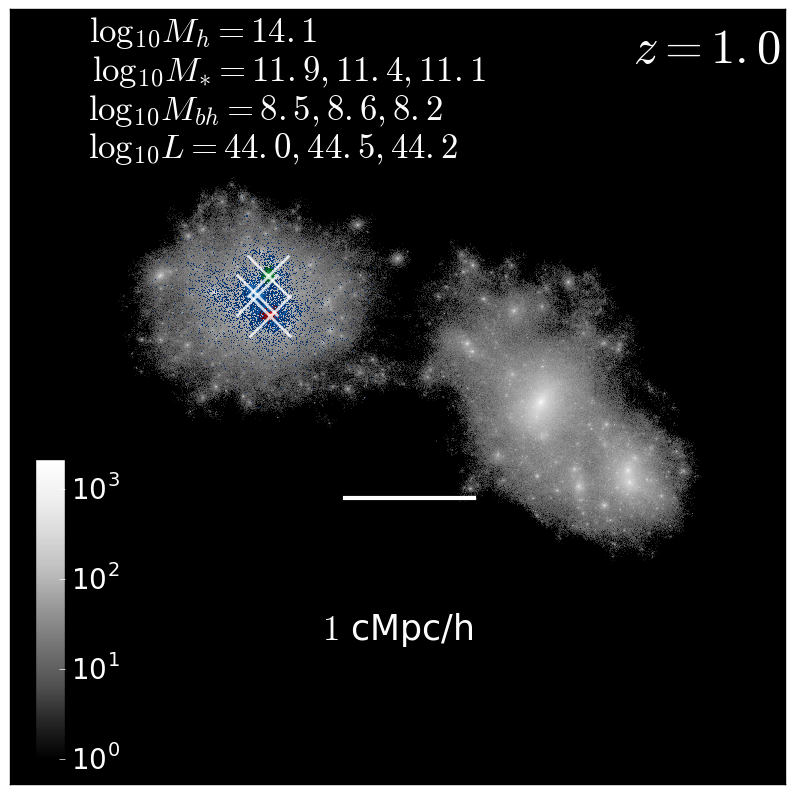}&
\includegraphics[width=5cm]{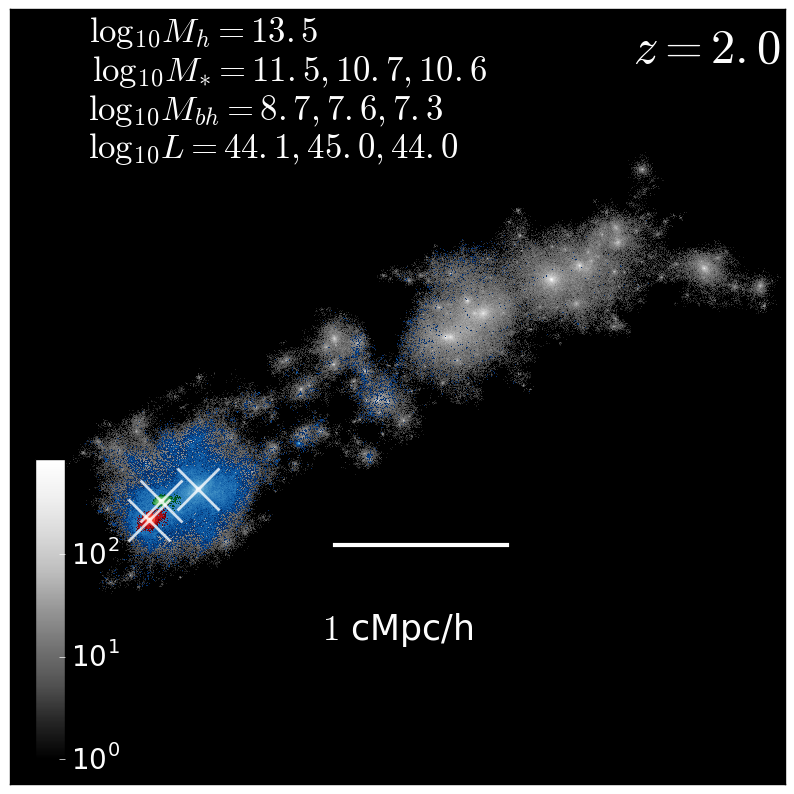}&
\includegraphics[width=5cm]{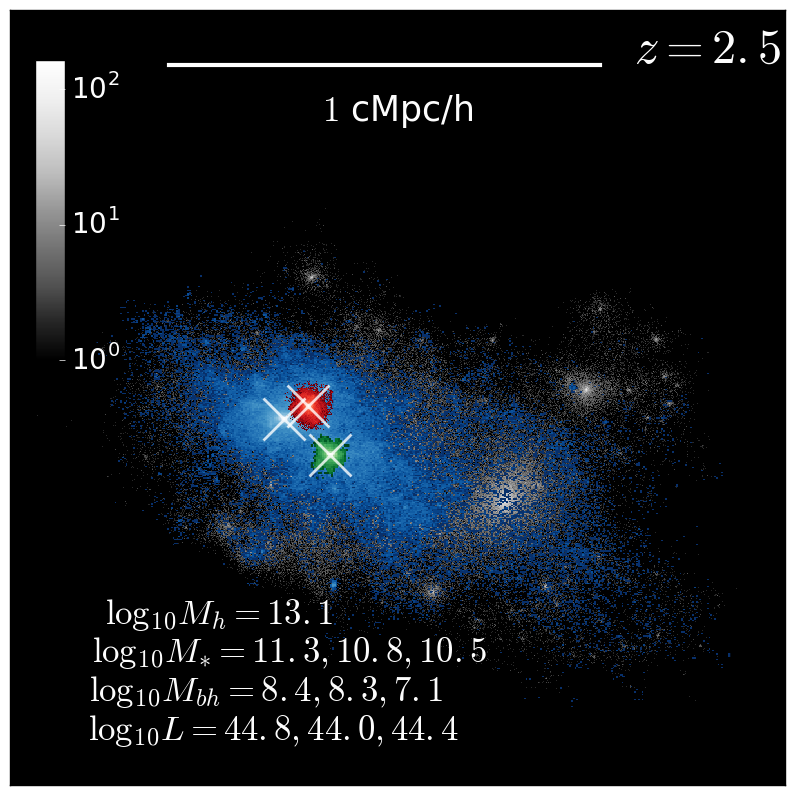}\\
\includegraphics[width=5cm]{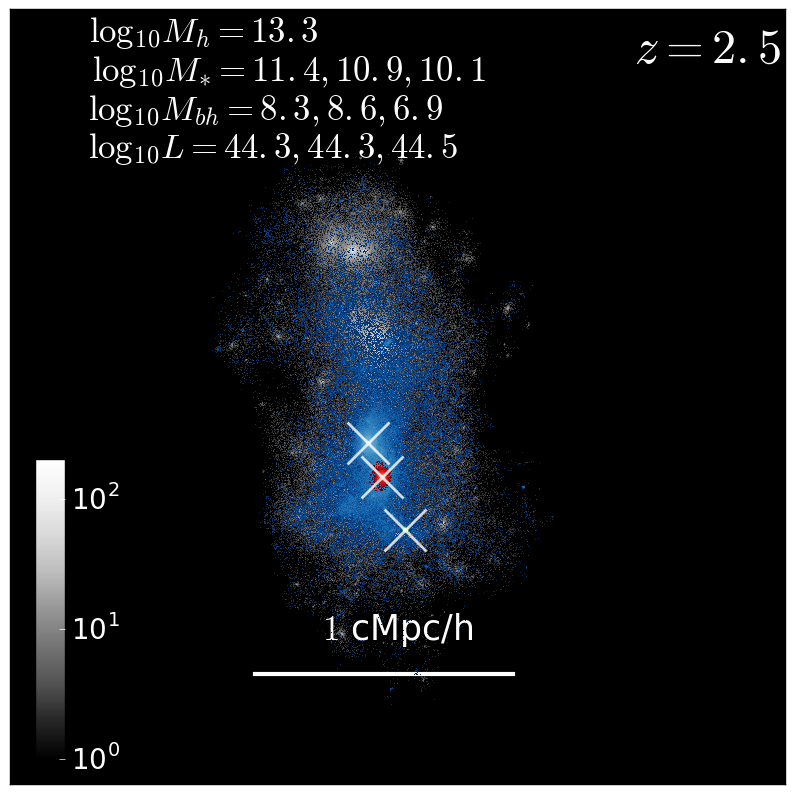}&
\includegraphics[width=5cm]{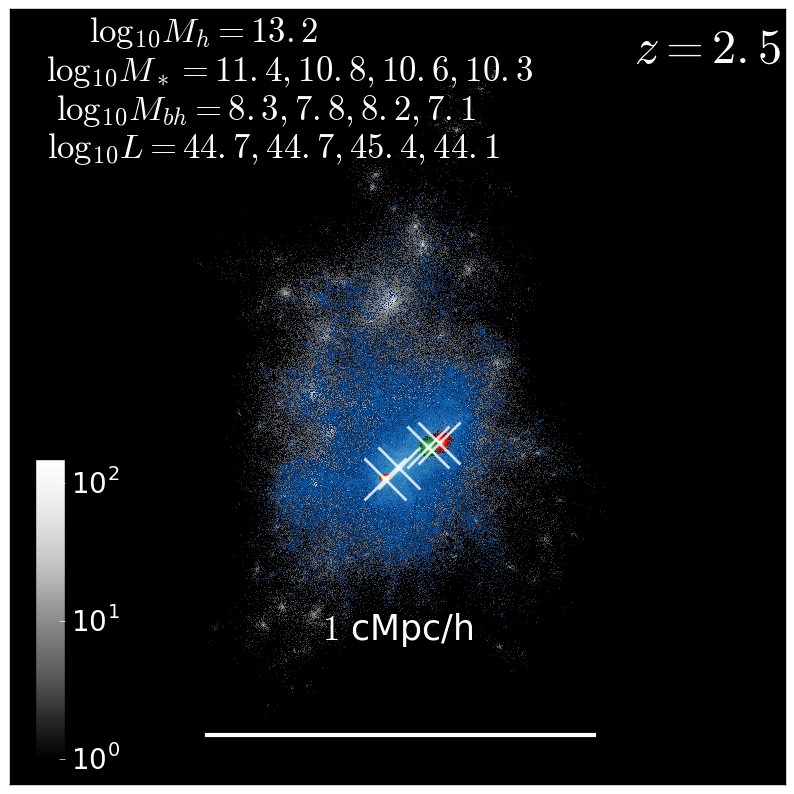}&
\includegraphics[width=5cm]{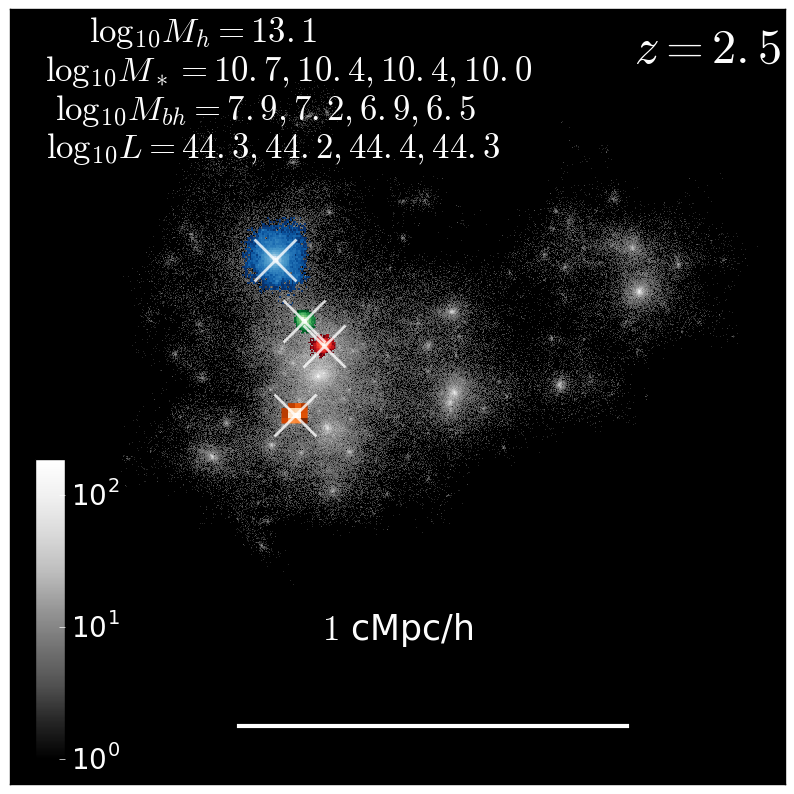}\\
\end{tabular}
\caption{The crosses indicate the 2D projected positions of the simulated quasar triples and quadruples with $d_{\mathrm{max}}=0.7~\mathrm{cMpc}/h$ and $L_{\mathrm{bol}}>10^{44}~\mathrm{ergs/sec}$. Grey 2-D density histograms represent the host dark matter haloes and colored histograms depict the host galaxies. The blue histogram corresponds to the central galaxy and the red, green, and orange histograms represent the satellite galaxies. Consult the legend for the various properties of the quasars and their hosts. $M_h$, $M_{*}$ and $M_{bh}$ are halo masses, stellar masses and black-hole masses in units of $M_{\odot}/h$. $L$ is the bolometric luminosity in the units of $\mathrm{ergs/sec}$. The horizontal white line represents a length scale of $1~\mathrm{cMpc}/h$. 
}
\label{quasar_triples}
\end{figure*}

\begin{figure*}

\includegraphics[width=\textwidth]{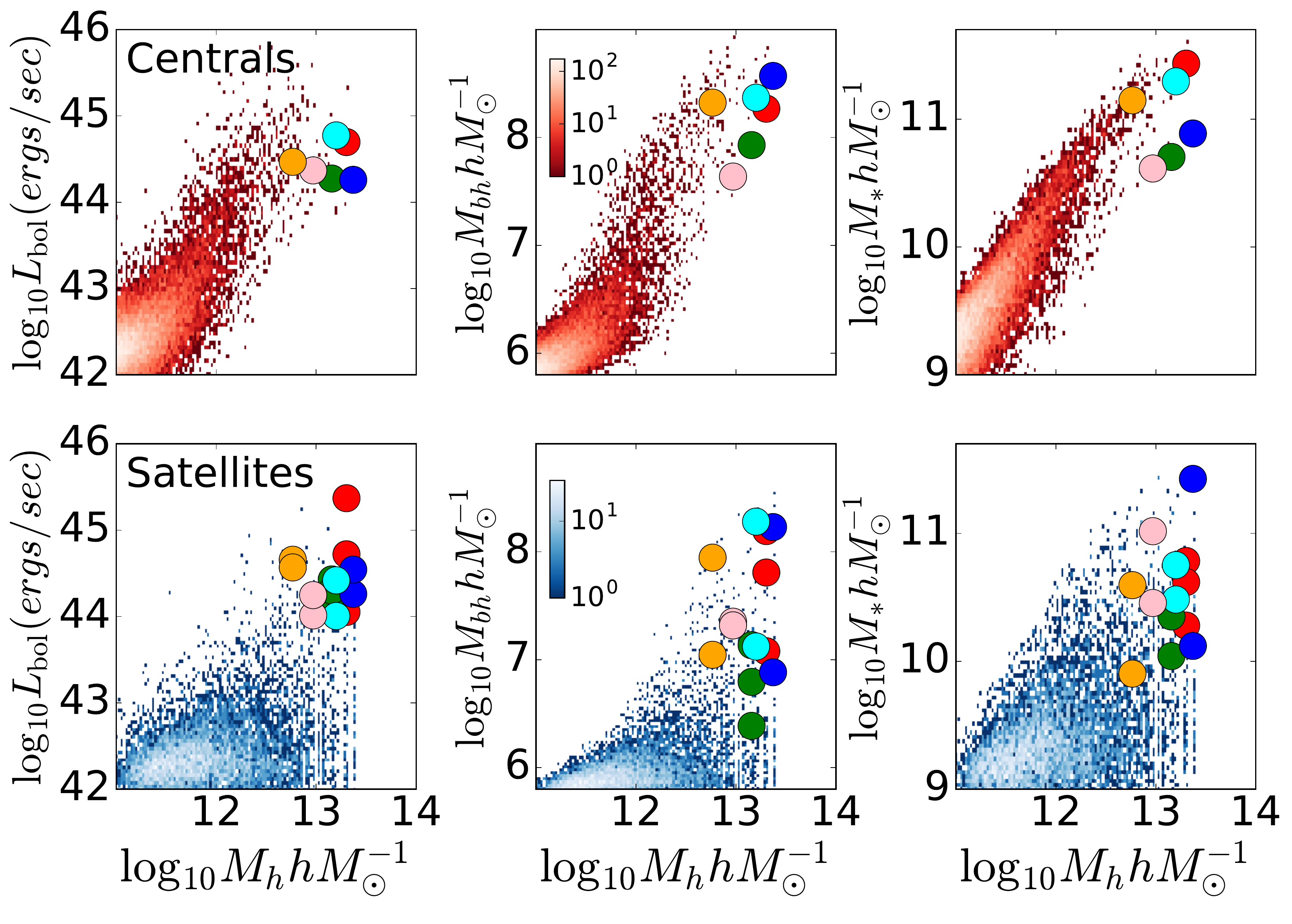}\\

\caption{Filled circles show the host halo masses, host galaxy masses,  black hole masses and luminosities of various quasar triples and quadruples with $L_{\mathrm{bol}}>10^{44}~\mathrm{ergs/sec}$ and $d_{\mathrm{max}}=0.7~\mathrm{cMpc}/h$ identified at $z=2.5$. The underlying histograms show the scaling relations between these properties for the overall AGN population at $z=2.5$. The blue histograms show satellite AGNs and red histograms show central AGNs.
}
\label{host_relations}
\end{figure*}

Until now, we have focused on the abundances of AGN systems. Hereafter, we shall investigate the physical properties and environment of the most luminous AGN~(quasar) systems in MBII. We will also study the growth histories of quasar systems to elucidate the physical mechanisms involved in their formation.  

We focus on systems where each member has a luminosity~$L_{\mathrm{bol}}>10^{44}~\mathrm{ergs/sec}$. Figure \ref{quasar_triples} shows images of these systems, along with their host galaxies (colored histograms) and host haloes (shown as grey histograms). The host halo masses for the systems are $\sim 10^{13}~M_{\odot}/h$. We note that these are amongst the most massive haloes in the simulation at the corresponding redshifts, in support of observational inferences \citep{2013MNRAS.431.1019F,2015Sci...348..779H} about these systems being progenitors of massive clusters; this is also consistently true for the simulated triples at $z=2$ and $z=1$, which live in $\sim 10^{13.5}~M_{\odot}/h$ and $\sim 10^{14}~M_{\odot}/h$ haloes respectively. Figure \ref{quasar_triples} shows that these systems have a very rich substructure i.e. a number of locally dense regions and complex morphologies, indicating possible occurrence of recent mergers. Within a halo, the member quasars live in different host galaxies (there are no $L_{\mathrm{bol}}>10^{44}~\mathrm{ergs/sec}$ quasar systems in MBII within the same galaxy). In particular, one member is hosted by the central~(most massive) galaxy, and the other members are hosted by satellite galaxies. The central galaxies are illustrated in Figure \ref{quasar_triples} as blue histograms. The satellite galaxies are illustrated as red, green and orange histograms. The host galaxies have stellar masses in the range $10^{10}~\lesssim M_* \lesssim~10^{12}~M_{\odot}/h$. Their black hole masses range from $\sim10^{6.5}$ to $\sim10^{9}~M_{\odot}/h$. 

Figure \ref{host_relations} shows the relationship of the various AGN properties~(bolometric luminosity, black hole mass and host galaxy stellar mass) as a function of the host halo mass at $z=2.5$. The filled circles correspond to the quasar systems whereas the histograms show the overall population of simulated AGNs. The top and bottom panels correspond to central and satellite AGNs respectively. Circles of the same color correspond to the central~(top panels) and satellite~(bottom panels) member quasars within the same system. In the AGN luminosity vs. halo mass  plane~(lower-left panel of Figure \ref{host_relations}), we can see that the satellite quasars have luminosities $\gtrsim10-100$ times higher than the typical population of satellite AGNs residing in the respective host haloes~($M_h\gtrsim10^{13}~M_{\odot}/h$). The black-hole mass vs. halo mass plane (lower-middle panel) shows that a handful of satellite quasars also have black hole masses $\gtrsim10-100$ times higher than a typical satellite. However, the stellar mass vs. halo mass plane~(lower right panel) shows that the host galaxies are more or less representative of the typical population of satellite AGN hosts. Overall, this hints at the possibility that the satellite quasars experienced a recent increase in their AGN activity. In the next subsection, we shall look at the growth histories of these AGNs and demonstrate that the increase in activity amongst satellite AGNs can be explained by recent major mergers amongst their progenitor host haloes.

%The stellar and black hole masses of the satellite galaxies~(middle and right panels) are $\gtrsim10$ times higher than average~(i.e. $M_*\sim10^{9}~M_{\odot}/h$ and $M_{bh}\sim10^{6}~M_{\odot}/h$).  

%Additionally, in the middle and right panels, we also see a handful of member quasars with luminosities $\gtrsim10$ times higher compared to expectations from their stellar and black hole masses. This suggests the possibility of events~(likely halo mergers) that triggered AGN activity in these objects causing a sudden increase in their luminosity.  

\label{individual_systems_sec}
\subsection{Growth Histories}
\label{growth_histories_sec}
\begin{figure}
\includegraphics[width=8cm]{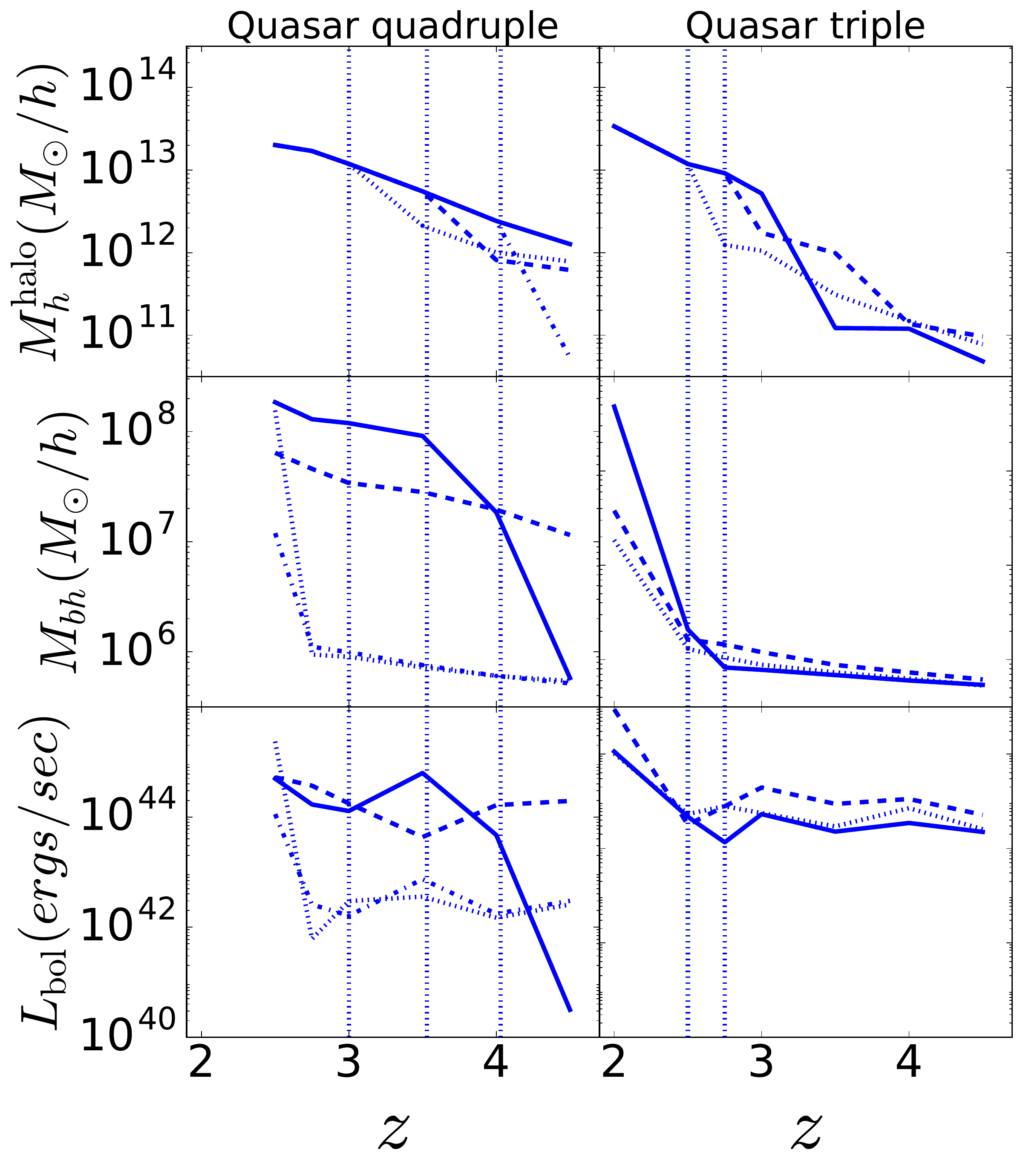}
\caption{Growth history of an example quasar quadruple (left panels) and triple (right panels) shown in Figure \ref{quasar_triples}. The panels from top to bottom correspond to the host halo mass, black hole mass and bolometric luminosity of the AGNs. The different line styles correspond to the different members of the system. Vertical lines of corresponding color show the time steps at which mergers of the host haloes occurred.}
\label{history_fig}
\end{figure}
We now study the growth histories of the simulated triples and quadruples. Figure \ref{history_fig} shows the growth history of an example triple (right panel) and quadruple (left panel). The different line styles~(solid, dashed, dot-dashed and dotted) correspond to different members of the system. The lowest redshift plotted for each line corresponds to the snapshot at which the objects form quasar systems with $L_{\mathrm{bol}}>10^{44}~\mathrm{ergs/sec}$.   

Let us first focus on the quadruple~(left-hand panel). The halo mass evolution~(top panel) shows that the progenitors start out in four different haloes that experience a sequence of three mergers between $z=4$ and $z=2.5$~(shown as dotted vertical lines). Each merger is followed by an increase in AGN activity amongst one or both of the participating progenitors. For example, for the dotted line, the merger occured at $z=2.75$ after which the luminosity increased from $L_{\mathrm{bol}}<10^{42}~\mathrm{ergs/sec}$ at $z=2.75$  to $L_{\mathrm{bol}}>10^{45}~\mathrm{ergs/sec}$ at $z=2.5$. Concurrently, the black hole mass increased from $M_{bh}\lesssim10^6~M_{\odot}/h$ at $z=2.75$ to $M_{bh}\gtrsim10^8~M_{\odot}/h$ at $z=2.5$. 

We report similar conclusions for the growth history of the triples (right panel of Figure \ref{history_fig} shows an example). The formation of each triple is associated with a sequence of two mergers of the host haloes of progenitor AGNs. These mergers triggered AGN activity to switch on the `quasar' mode of the AGNs residing in these merging haloes. %A particularly striking example of the merger-driven quasar activity is that of the pink solid line. A halo merger occurring at $z=1.5$ was followed by an increase in the luminosity from $\sim 10^{38}~\mathrm{ergs/sec}$ to $\sim 10^{44}~\mathrm{ergs/sec}$. 

While Figure \ref{history_fig} shows only a couple of examples of quasar triples/ quadruples, our findings are true for all the systems shown in Figure \ref{quasar_triples}. In other words, all the multiple-quasar-systems in MBII originated from a sequence of multiple major mergers. These mergers not only bring together already active quasars, but also served as triggers of AGN activity transforming faint AGNs into luminous~($L>10^{44}~\mathrm{ergs/sec}$) quasars. Within MBII, we identified 5 members belonging to triple/ quadruple systems which were converted from faint AGNs~($L<10^{42}~\mathrm{ergs/sec}$) into luminous quasars~($L>10^{44}~\mathrm{ergs/sec}$) after a major merger. This merger driven activity was crucial to the formation of the quasar triples and quadruples. In fact, significant observational evidence of enhanced AGN activity in merged systems have also been reported in the last decade~\citep{2011ApJ...737..101L,2011ApJ...743....2S,2011MNRAS.418.2043E,2013MNRAS.435.3627E,2014AJ....148..137L,2014MNRAS.441.1297S,2017MNRAS.464.3882W,2018PASJ...70S..37G,2019MNRAS.487.2491E}. Furthermore, we see that for all the simulated systems, haloes which merge belong to the rare massive end of the halo mass function at the corresponding redshift. In our previous paper~\citep{2019MNRAS.485.2026B}, we showed that bright quasar pairs originate from mergers of rare massive haloes. Here, we further show that in order to form \textit{richer} quasar systems such as quasar triples, quadruples etc., we require a multiple sequence~(more than one) of mergers of rare massive haloes. These events are exceptionally infrequent, explaining the sparsity of these quasar systems.            
\section{Conclusions and Future work}
\label{conclusions}
In this work, we have studied systems of AGNs at $0\lesssim z\lesssim4$ and determine AGN multiplicity functions within the MBII simulation. We identified AGN systems at different scales by linking together AGNs within a certain distance. We find that to target gravitationally bound~(within the same halo) AGN systems in observations, the maximum comoving distance between member AGNs should be $\sim0.7~\mathrm{cMpc}/h$; this corresponds to maximum angular separations ranging from $\lesssim100~\mathrm{arcsec}$ at $z\sim0.6$ to $\lesssim30~\mathrm{arcsec}$ at $z\sim4$. Along the line of sight this corresponds to maximum velocity differences ranging from $\lesssim 150~\mathrm{km/sec}$ at $z\sim0.6$ to $\lesssim 200~\mathrm{km/sec}$ at $z\sim4$.%; however, peculiar velocity dispersion of AGNs tends to increase their redshift space separation. Therefore, we require a compensatory increase in the spectroscopic redshift difference to $\lesssim 5\times 10^{-3}$, so as to not miss objects that are gravitationally bound.

%Incidentally, we find that abundances of AGN \textit{projected} pairs in MBII are consistent with constraints from SDSS at $0.02\lesssim z \lesssim 0.15$; $\gtrsim85~\%$ of these simulated pairs are gravitationally bound. This motivates us to move further and use MBII as a tool to study the origin and abundance of higher order AGN systems~(triples/ quadruples).  

The multiplicity functions and their redshift evolution reveal that AGN systems~(pairs, triples, quadruples) in MBII are most abundant at $1.5\lesssim z \lesssim 2.5$ wherein there are $10^{-5}$- $10^{-6}~h^3\mathrm{Mpc^{-3}}$ quasar triples with $L_{\mathrm{bol}}\gtrsim10^{44}~\mathrm{ergs/sec}$. We find that the dependence of the multiplicity function as a function of $R$ can be well described by a power law with exponents ranging from $-3$ to $-6$. MBII directly probes AGN systems with magnitudes up to $g\sim24$ at $0.06\lesssim z\lesssim4$. We predict abundances of ~10 triples or quadruples per $\mathrm{deg}^2$ at $g < 24$ (depth of DESI imaging) and ~100 triples or quadruples per $\mathrm{deg}^2$ at $g < 26$ (depth of LSST imaging). To make predictions for $g<22$~(depth of eBOSS-CORE), we used HOD modeling; we predict a few $\times10^{-2}~\mathrm{deg}^{-2}$ AGN triples/quadruples at $0.9\lesssim z \lesssim 2.2$. Using a simple model to describe fiber collisions, we predict that $\sim20\%$ of the available triples or quadruples should be detectable after spectroscopic follow-up. 

Finally, in order to probe the possible physical origins of these systems, we select a few of the most luminous ($L_{\mathrm{bol}}>10^{44}~\mathrm{ergs/sec}$) quasar triples and quadruples and study their environmental properties and growth histories. We find that their host haloes are amongst the most massive in the simulation ($\sim10^{13}~M_{\odot}/h$ at $z=2.5$ and $\sim10^{14}~M_{\odot}/h$ at $z=1$). Within these haloes, the members of quasar triples and quadruples always reside in separate galaxies, implying that one quasar lives in a central galaxy and the remaining quasars live in satellite galaxies. Note however that the simulations tend to assume instantaneous merger between black holes; a more accurate modeling of black hole mergers will likely lead to higher number of AGN systems living within the same host galaxy. Such models will therefore also lead to enhanced multiplicity functions particularly at small~($\lesssim1~\mathrm{kpc}$) distances of separation. Furthermore, the gravitational recoil resulting from the merging process will influence the time evolution of the pair distances, and therefore may have a significant impact on the multiplicity functions. Future work shall investigate these aspects in detail. The galaxies have stellar masses $10^{10}\lesssim M_{*}\lesssim 10^{12}~M_{\odot}/h$ and black hole masses $10^{6}\lesssim M_{*}\lesssim 10^{9}~M_{\odot}/h$. Growth histories of these systems reveal that they were born out of a series of mergers~(two mergers for triples and three mergers for quadruples) between rare massive haloes. These mergers can not only produce close pairs from already active quasars, but can also trigger significant activity in low-luminosity AGNs, transforming them into luminous quasars. This explains the wide range of black hole massses ($10^{6.5}\lesssim M_{bh}\lesssim 10^9~M_{\odot}/h$) for these objects. Sequences of multiple~(two or more) mergers between the most massive haloes are exceedingly uncommon, explaining why these quasar systems are so rare. 

Future work(s) can include investigations of further limitations in the detectibility of these AGN systems in observations. For instance, the member AGNs of these systems will have variability and are not always active; it is therefore important to determine how much this will reduce the probability of simultaneous detection of all the members of the system, and its impact on the observed multiplicity functions. Other avenues include an investigation of the effect of black hole seed models on the multiplicity functions, as well as predictions of gravitational wave signals sourced from such systems.

\section*{Acknowledgements}
AKB, TD and ADM were supported by the National Science Foundation through grant number 1616168. TDM acknowledges funding from NSF
ACI-1614853, NSF AST-1517593, NASA ATP NNX17AK56G and NASA ATP 17-0123. The \texttt{BLUETIDES} simulation was run on the BlueWaters facility at the National Center for Supercomputing Applications. ADM also acknowledges support through the U.S. Department of Energy, Office of Science, Office of High Energy Physics, under Award Number DE-SC0019022. 
\bibliography{references}

\appendix

\section{Modeling of AGN HODs using Conditional Luminosity Function~(CLF) formalism}
\label{mean_occupations_sec}
We use AGN HODs which were derived in our previous work \citep{2019MNRAS.485.2026B} from small scale clustering constraints~\citep{2017MNRAS.468...77E} at $0.6\lesssim z \lesssim 2$. In particular, we used the Conditional Luminosity function~(CLF) model~\citep{2006MNRAS.365..842C} to parametrize the HODs. Under this model, for a given sample of AGNs with a luminosity threshold $L_{\mathrm{min}}$, the first order moment  $\left<N(L>L_{\mathrm{min}},M_h)\right>$ is given by   
\begin{equation}
\left<N(L>L_{\mathrm{min}},M_h)\right>=\int_{L_{\mathrm{min}}}^{\infty}~\tilde{\Phi}(L',M_h)~d\log_{10}L' 
\label{CLF_to_HOD}
\end{equation}
where $\tilde{\Phi}(L,M_h)$ is the Conditional luminosity function~(CLF) defined as the distribution of AGN bolometric luminosities within haloes of masses $M_h$ to $M_h+d M_h$, such that  
\begin{equation}
\int~\tilde{\Phi}(L,M_h)~\frac{dn}{dM_h}~dM_h=\Phi(L)  
\end{equation}
where $\Phi(L)$ is the total quasar bolometric luminosity function. 

The CLF can be separated into contributions from centrals~($\tilde{\Phi}_{\mathrm{cen}}(L,M_h)$) and satellites~($\tilde{\Phi}_{\mathrm{sat}}(L,M_h)$)  
\begin{equation}
    \tilde{\Phi}(L,M_h)=\tilde{\Phi}_{\mathrm{cen}}(L,M_h)+\tilde{\Phi}_{\mathrm{sat}}(L,M_h).
\end{equation}
The CLF for central quasars can be parametrized as a log normal distribution
\begin{eqnarray}
\nonumber \tilde{\Phi}_{\mathrm{cen}}(L,M_h)= Q_{\mathrm{cen}}~\frac{1}{2 \pi \sigma_{\mathrm{cen}}} \exp{\left(-\frac{(\log_{10}{L}-\log_{10}{L^*_{\mathrm{cen}}})^2}{\sigma_{\mathrm{cen}}^{2}}\right)} ,
\label{central_CLF_eqn}
\end{eqnarray}
where $L^*_{\mathrm{cen}}$ is the average central AGN luminosity, $\sigma_{\mathrm{cen}}$ is the width of the log-normal distribution. $Q_{\mathrm{cen}}$ is the normalization of the central AGN. The CLFs for satellite quasars can be parametrized as a Schechter distribution
\begin{eqnarray}
\nonumber \tilde{\Phi}_{\mathrm{sat}}(L,M_h)= Q_{\mathrm{sat}} \left(\frac{L}{L^*_{\mathrm{sat}}}\right)^{\alpha_{\mathrm{sat}}}\exp{\left(-\frac{L}{L^*_{\mathrm{sat}}}\right)}
\label{satellite_CLF_eqn}
\end{eqnarray}
$L^{*}_{\mathrm{sat}}$ measures the most luminous satellite for a given halo of mass $M_h$, $\alpha_{\mathrm{sat}}$ is the faint end slope of the distribution, and $Q_{\mathrm{sat}}$ is the normalization.

\begin{figure*}
\includegraphics[width=\textwidth]{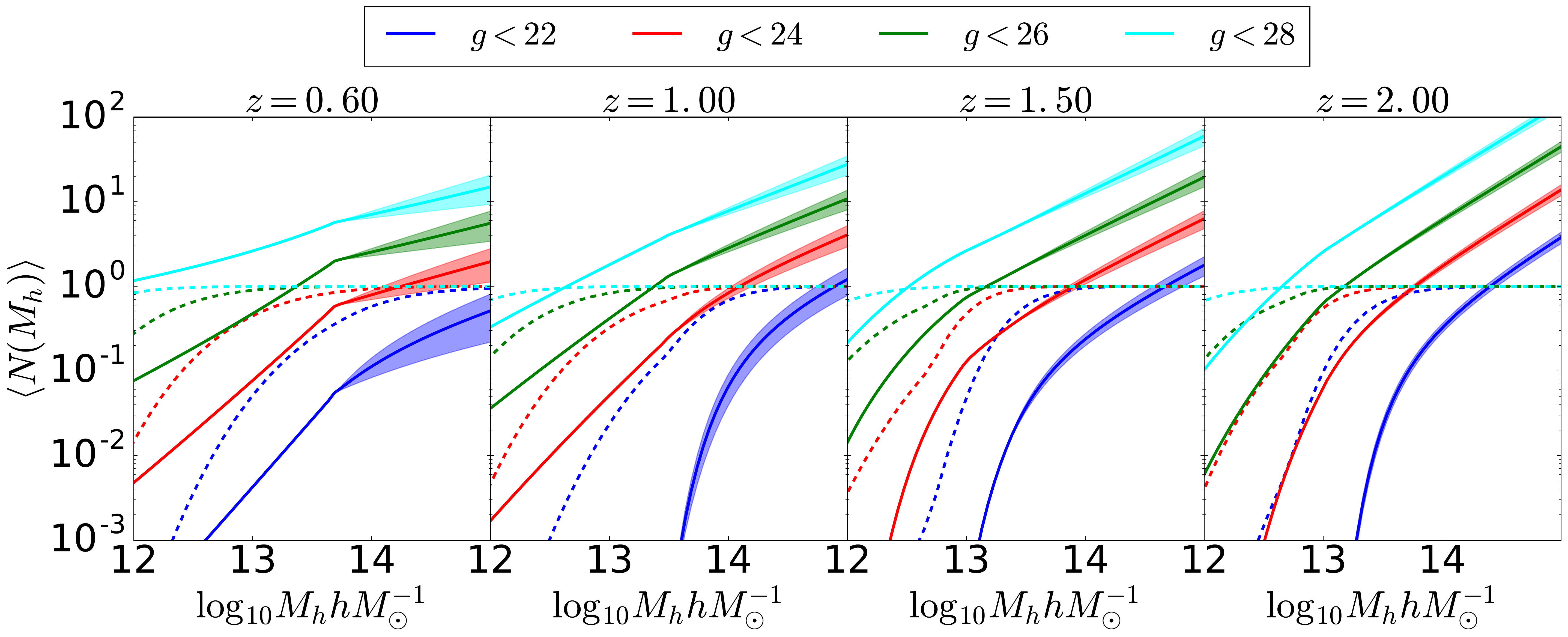}
\caption{Mean occupations of satellite~(solid lines) and central AGNs~(dashed lines) as a function of halo mass over a redshift range of $0.6\lesssim z\lesssim2$. The shaded regions correspond to uncertainties in the small scale~($\sim 25~\mathrm{kpc}/h$ ) clustering constraints in \protect\cite{2017MNRAS.468...77E}.}
\label{HODs_ref}
\end{figure*}

In \cite{2019MNRAS.485.2026B}, we used the simulation predictions and observed measurements of the luminosity function and the small scale clustering to constrain the parameters $Q_{\mathrm{cen}},\sigma_{\mathrm{cen}}, L^*_{\mathrm{cen}},Q_{\mathrm{sat}},\alpha_{\mathrm{sat}},L^*_{\mathrm{sat}}$ to determine $\tilde{\Phi}(L,M_h)$. We then used Eq.~\ref{CLF_to_HOD} to finally obtain the 1st order moments $\left<N\right>$ of central and satellite AGN HODs. Figure \ref{HODs_ref} shows the 1st order moments of the resulting mean occupations of central and satellite AGNs at $z=0.6,1,1.5,2$. In order to arrive the HODs~($P(N|M_h)$ as defined in Section \ref{HOD_sec}) from these moments, the satellite occupations are assumed to be Poisson. From $P(N|M_h)$, one can then derive the multiplicity function using Eqs.~(\ref{surface_from_volume}-\ref{HOD_model_eqn}).

%From the mean occupations, one can infer halo masses that contain \textit{bound} AGN triples and quadruples for a given redshift and threshold magnitude.} 

\end{document}